\documentclass[pra, showpacs, preprint, superscriptaddress]{revtex4}
\usepackage{graphicx}    
\usepackage{bm}          

\usepackage{color}
\usepackage{CJK}
\begin{document}
\title{Nonlinear Fluid Simulation Study of Stimulated Raman and Brillouin Scatterings in Shock Ignition}

\author{L. Hao\footnote{Permanent Address: Institute of Applied Physics and Computational Mathematics, Beijing, 100094, China}}
\affiliation{Department of Mechanical Engineering and Laboratory for
Laser Energetics, University of Rochester, Rochester, New York
14627, USA}

\author{{R. Yan}\footnote{Email: ruiyan@ustc.edu.cn}}
\affiliation{Department of Mechanical Engineering and Laboratory for
Laser Energetics, University of Rochester, Rochester, New York
14627, USA}
\affiliation{University of Science and Technology of China, Hefei,
Anhui, 230026, China}

\author{J. Li}
\affiliation{Department of Mechanical Engineering and Laboratory for
Laser Energetics, University of Rochester, Rochester, New York
14627, USA}

\author{W. D. Liu}
\affiliation{Department of Mechanical Engineering and Laboratory for Laser Energetics, University of Rochester,
Rochester, New York 14627, USA}

\author{C. Ren\footnote{Email: chuang.ren@rochester.edu}}

\affiliation{Department of Mechanical Engineering and Laboratory for Laser Energetics, University of Rochester,
Rochester, New York 14627, USA}
\affiliation{Department of Physics and Astronomy, University of Rochester, Rochester, New York 14627, USA}

\date{\today}

\begin{abstract}
We developed a new nonlinear fluid laser-plasma-instability code
({\it FLAME}) using a multi-fluid plasma model combined with full
electromagnetic wave equations. The completed one-dimensional (1D)
version of {\it FLAME} was used to study laser-plasma instabilities
in shock ignition. The simulations results showed that absolute SRS
modes growing near the quarter-critical surface were saturated by
Langmuir-wave Decay Instabilities (LDI) and pump depletion. The
ion-acoustic waves from LDI acted as seeds of Stimulated Brillouin
Scattering (SBS), which displayed a bursting pattern and caused
strong pump depletion. Re-scattering of SRS at the 1/16th-critical
surface was also observed in a high temperature case. These results
largely agreed with corresponding Particle-in-Cell simulations.

\end{abstract}

\maketitle

\section{Introduction}

Laser-plasma instabilities (LPI) in inertial confinement fusion
(ICF) include both convective and absolute modes. While absolute
modes can grow to large amplitudes locally and are saturated by
nonlinear effects, convective modes growth is limited by density
gradients and their saturation amplitudes are the product of the
seed level and the so-called Rosenbluth gain \cite{Rosenbluth72}. In
recent particle-in-cell (PIC) simulations of LPI in a new ignition
scheme, shock ignition \cite{Betti07}, convective and absolute modes
were found both important. Laser reflectivity was found largely due
to convective stimulated Brillouin scattering (SBS) below the
quarter-critical surface \cite{Riconda11,Weber12,Weber15,Hao16}. Hot
electrons were generated by absolute two plasmon decay (TPD) and
stimulated Raman scattering (SRS) modes near the quarter-critical
surface \cite{Riconda11,Weber12,Weber15,Yan14}. Convective SRS was
also found below the quarter-critical surface
\cite{Riconda11,Weber12} accompanied by significant changes in the
electron distribution function and corresponding Landau damping
\cite{Riconda11}. However it was not clear how the convective SRS
can grow to such large amplitudes to alter the electron distribution
function. In a recent 1D PIC simulations with a long density scale
\cite{Hao16}, convective SRS was also found to contribute to the
laser reflectivity at a higher level than that from fluid
calculation based on a Thomson scattering seed level model
\cite{Berger89} and the the Rosenbluth gain. It was found that the
seed levels for both convective SBS and SRS in the PIC simulations
were 4-7 orders of magnitude higher than the thermal noise
\cite{Hao16}. However, the fluid calculation in \cite{Hao16} with a
ray-based steady-state code HLIP \cite{Hao14} did not include the
effects of the absolute SRS near the quarter-critical surface. It
was not clear whether the high SRS reflectivity in the PIC
simulations was entirely due the high seed levels. It was also not
clear whether the large SBS reflectivity in the PIC simulations was
influenced by the high seed levels.

To bridge the gap between full PIC simulations and steady-state
fluid simulation on laser-plasma instabilities in ICF, especially in
shock ignition, we developed a new nonlinear \textbf{FL}uid code of
l\textbf{A}ser-plas\textbf{M}a instabiliti\textbf{E}s ({\it FLAME}).
To make the physics as close to PIC simulations as possible, we use
the full wave equations for vector potentials of light without
envelope approximations. Thus no particular mode would be precluded.
The plasma is represented by an electron species and multiple ion
species. Landau damping for both electrons and ions is evaluated in
the spectral-space and included in the momentum equations for each
species. Our model is in principle multi-dimensional (see the
Appendix for detail) and can allow all three major LPI modes (SRS,
SBS, and TPD). In this paper we report the implementation and
benchmark of a one-dimension (1D) parallel version of {\it FLAME}
and some simulation results on SRS and SBS in the shock ignition
regime. Similar to the PIC simulations \cite{Hao16}, absolute SRS
near the quarter-critical surface were observed to saturate via
Langmuir wave decay instability (LDI) \cite{Karttunen81} and pump
depletion. Re-scattering of the absolute SRS was also observed when
the electron temperature was high. Laser reflectivity due to
convective SRS was much smaller than in the PIC simulations, due to
the seed level difference. The SBS reflectivity was similar to that
in the PIC simulations and is the main cause for pump depletion.
Through a contrasting simulation, we concluded that some ion
acoustic wave modes generated in LDI acted as seeds for SBS,
resulting bursting patterns in both SBS and transmitted light.

\section{Implementation of the 1D Version of {\it \textbf{FLAME}}}
Details of our physics model are presented in the Appendix. In 1D,
Eqs. (\ref{first:eq})-(\ref{last:eq}) in the Appendix can be written
in a dimensionless form as

\begin{eqnarray}\label{e15}
\left( {\frac{{{\partial ^2}}}{{\partial {t^2}}} + {\nu
_{c0}}\frac{\partial }{{\partial t}} - \frac{{{\partial
^2}}}{{\partial {x^2}}} + {n_A}} \right){A_0} =  - {n_L}{A_1},
\end{eqnarray}
\begin{eqnarray}\label{e16}
\left( {\frac{{{\partial ^2}}}{{\partial {t^2}}} + {\nu
_{c1}}\frac{\partial }{{\partial t}} - \frac{{{\partial
^2}}}{{\partial {x^2}}} + {n_A}} \right){A_1} =  - {n_L}{A_0},
\end{eqnarray}
\begin{eqnarray}\label{e17}
\frac{{\partial {n_L}}}{{\partial t}} =  - \frac{\partial
}{{\partial x}}({n_A}{u_{Lx}} + {n_L}{u_{Ax}}),
\end{eqnarray}
\begin{eqnarray}\label{e18}
\frac{{\partial {u_{Lx}}}}{{\partial t}} &+& ({\nu _{ce}} + 2{\nu
_{Le}}){u_{Lx}}\\\nonumber &=& \frac{\partial }{{\partial x}}({\phi
_L} - {A_0}{A_1} - {u_{Ax}}{u_{Lx}})\\\nonumber &&-
\frac{{{3}}}{{{n_A}}}\frac{\partial }{{\partial x}}({T_e}{n_L})+S_L,
\end{eqnarray}
\begin{eqnarray}\label{e19}
\frac{{{\partial ^2}}}{{\partial {x^2}}}{\phi _L} = {n_L},
\end{eqnarray}
\begin{eqnarray}\label{e20}
{n_A} = \sum\limits_j {{Z_j}{n_j}} ,
\end{eqnarray}
\begin{eqnarray}\label{e21}
{n_A}{u_{Ax}} = \sum\limits_j {{Z_j}{n_j}} {u_{jx}} ,
\end{eqnarray}
\begin{eqnarray}\label{e22}
\frac{{\partial {n_j}}}{{\partial t}} =  - \frac{\partial
}{{\partial x}}({n_j}{u_{jx}}),
\end{eqnarray}
\begin{eqnarray}\label{e23}
\frac{{\partial {u_{jx}}}}{{\partial t}} &+& ({\nu _{cj}} + 2{\nu
_{Lj}}){u_{jx}}+\frac{1}{2}\frac{\partial }{{\partial
x}}u_{jx}^2\\\nonumber &=& -
\frac{{{Z_j}}}{{{m_j}{n_A}}}\frac{\partial }{{\partial
x}}({T_e}{n_A})- \frac{{{3}}}{{{m_j}{n_j}}}\frac{\partial
}{{\partial x}}({T_j}{n_j})\\\nonumber &&
-\frac{1}{2}\frac{{{Z_j}}}{{{m_j}}}\frac{\partial }{{\partial
x}}(u_{Lx}^2 + A_0^2 + A_1^2)+S_j.
\end{eqnarray}
Here, the normalized vector potential $A_0$ represents the incident
laser and all scattered light that is not primary SRS, including SBS
and the re-scattering of SRS, and $A_1$ represents the scattered
light from the primary SRS. The normalized electron density
$n_{A,L}$ and velocity $u_{A,Lx}$ represent the electron response on
the ion acoustic scale ({\it A}) and Langmuir wave scale ({\it L}).
The normalized ion density, velocity, mass, and charge state for
species $j$ are $n_j,u_{jx},m_j$ and $Z_j$, respectively. Both the
collisional ($\nu_c$) and Landau ($\nu_{L}$) damping are included.
The normalized units of density, time, length, velocity, mass,
temperature, vector potential and static electric potential are
$n_c$, $1/\omega_0$, $c/\omega_0$, $c$, $m_e$, $m_ec^2$, $m_ec^2/e$
and $m_ec^2/e$ respectively, where $n_c$ is the critical density and
$\omega_0$ is the frequency of the incident laser. $S_L$ and $S_j$
are random fluctuating source terms with the magnitudes
approximating to thermal noise. The 1D model was implemented in four
parts including the electromagnetic (EM) wave solver, the electron
response solver, the ion response solver, and the Landau damping
solver.

In the EM wave solver, Eqs. (\ref{e15}) and (\ref{e16}) were
discretized by the central difference scheme. Laser was launched
from an antenna at the left boundary of the physical domain. The
perfectly matched layer (PML) technique \cite{Berenger94} was used
as the absorption boundary conditions at both sides of the physical
domain.

For the electron response solver, Eq. (\ref{e18}) was separated into
\begin{eqnarray}\label{e25}
\frac{{\partial {u_{Lx}}}}{{\partial t}} + \nu _{ce}{u_{Lx}} &=&
\frac{\partial }{{\partial x}}({\phi _L} - {A_0}{A_1} -
{u_{Ax}}{u_{Lx}})\\\nonumber &&- \frac{{{3}}}{{{n_A}}}\frac{\partial
}{{\partial x}}({T_e}{n_L})+S_L,
\end{eqnarray}
\begin{eqnarray}\label{e26}
\frac{{\partial {u_{Lx}}}}{{\partial t}} + 2{\nu _{Le}}{u_{Lx}} = 0,
\end{eqnarray}
based on the time-splitting method \cite{Strang68}. Equations
(\ref{e17}) and (\ref{e25}) were solved together using a leapfrog
scheme. Equation (\ref{e26}) was solved in a Landau damping solver
described later in this section. The Poisson equation (\ref{e19})
was solved as a tri-diagonal matrix equation in the entire
simulation box.

Similarly, for the ion response solver, Eq. (\ref{e23}) was separated into
\begin{eqnarray}\label{e29}
\frac{{\partial {u_{jx}}}}{{\partial t}} &=&  -
\frac{1}{2}\frac{\partial }{{\partial x}}u_{jx}^2\\\nonumber && -
\frac{{{Z_j}}}{{{m_j}{n_A}}}\frac{\partial }{{\partial
x}}({T_e}{n_A}) - \frac{{{3}}}{{{m_j}{n_j}}}\frac{\partial
}{{\partial x}}({T_j}{n_j})\\\nonumber &&
-\frac{1}{2}\frac{{{Z_j}}}{{{A_j}{m_p}}}\frac{\partial }{{\partial
x}}(u_{Lx}^2 + A_0^2 + A_1^2)+S_j,
\end{eqnarray}
\begin{eqnarray}\label{e30}
\frac{{\partial {u_{jx}}}}{{\partial t}} + ({\nu _{cj}} + 2{\nu
_{Lj}}){u_{jx}} =0,
\end{eqnarray}
based on the time-splitting method. Equations (\ref{e30}) was also
solved in the Landau damping solver. Since $u_{jx}$ includes both
the background flow velocity and the perturbation components that
are the ion-acoustic waves, the collisional damping ($\nu_{cj}$) was
included in Eqs. (\ref{e30}), which was solved in the spectral
space, to apply only on the modes with nonzero wavenumbers. For each
time step, after calculating Eqs. (\ref{e20}) and (\ref{e21})
locally, Eqs. (\ref{e22}) and (\ref{e29}) were solved together using
the leapfrog scheme in both the physics domain and the PML layers
with the boundary condition $\partial^2 n_j/\partial x^2=0$ and
$\partial^2 u_{jx}/\partial x^2=0$.

\begin{figure*}[htb!]
(a)
\includegraphics[height=0.3\textwidth,width=0.35\textwidth,angle=0]{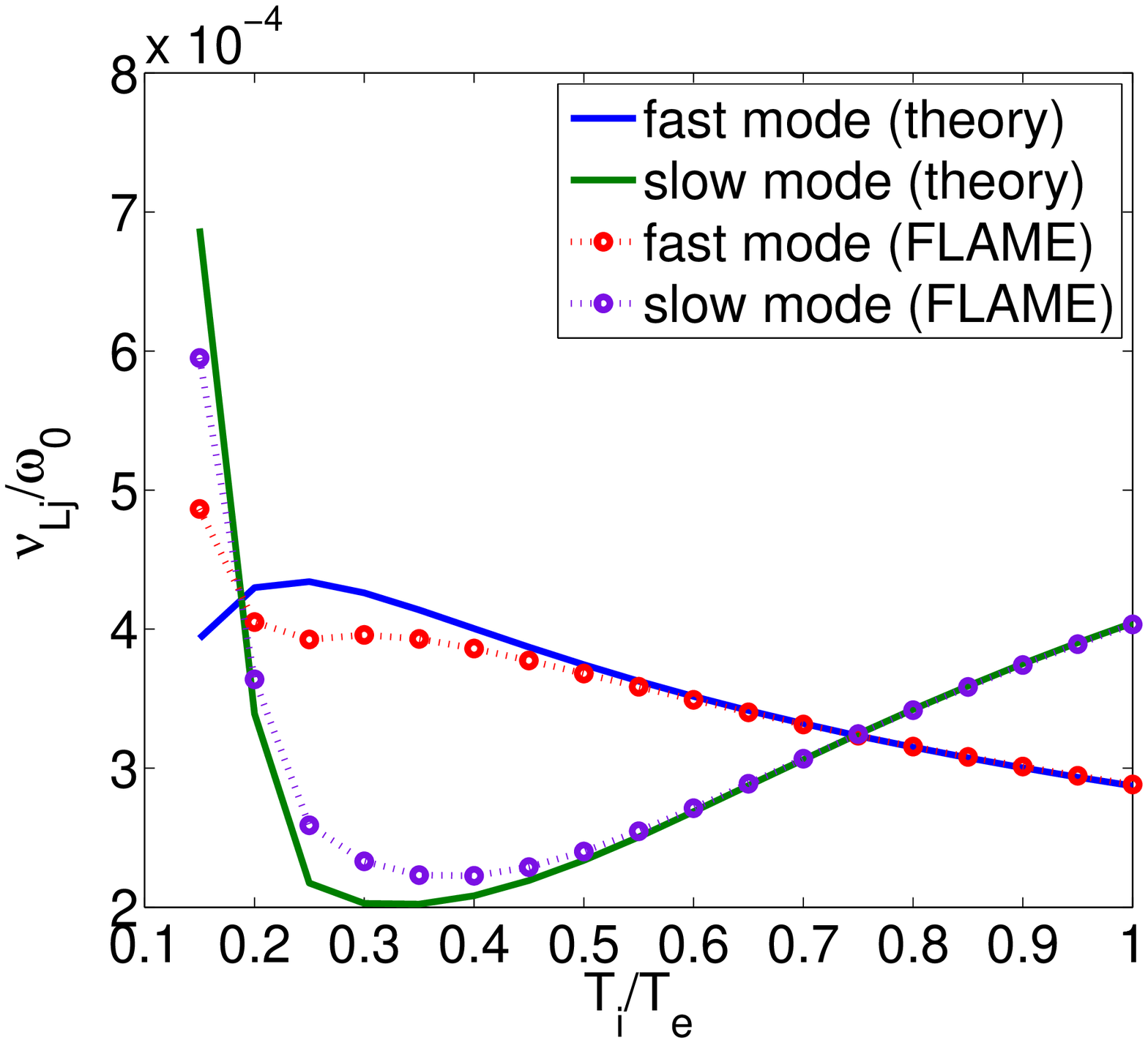}
(b)
\includegraphics[height=0.3\textwidth,width=0.35\textwidth,angle=0]{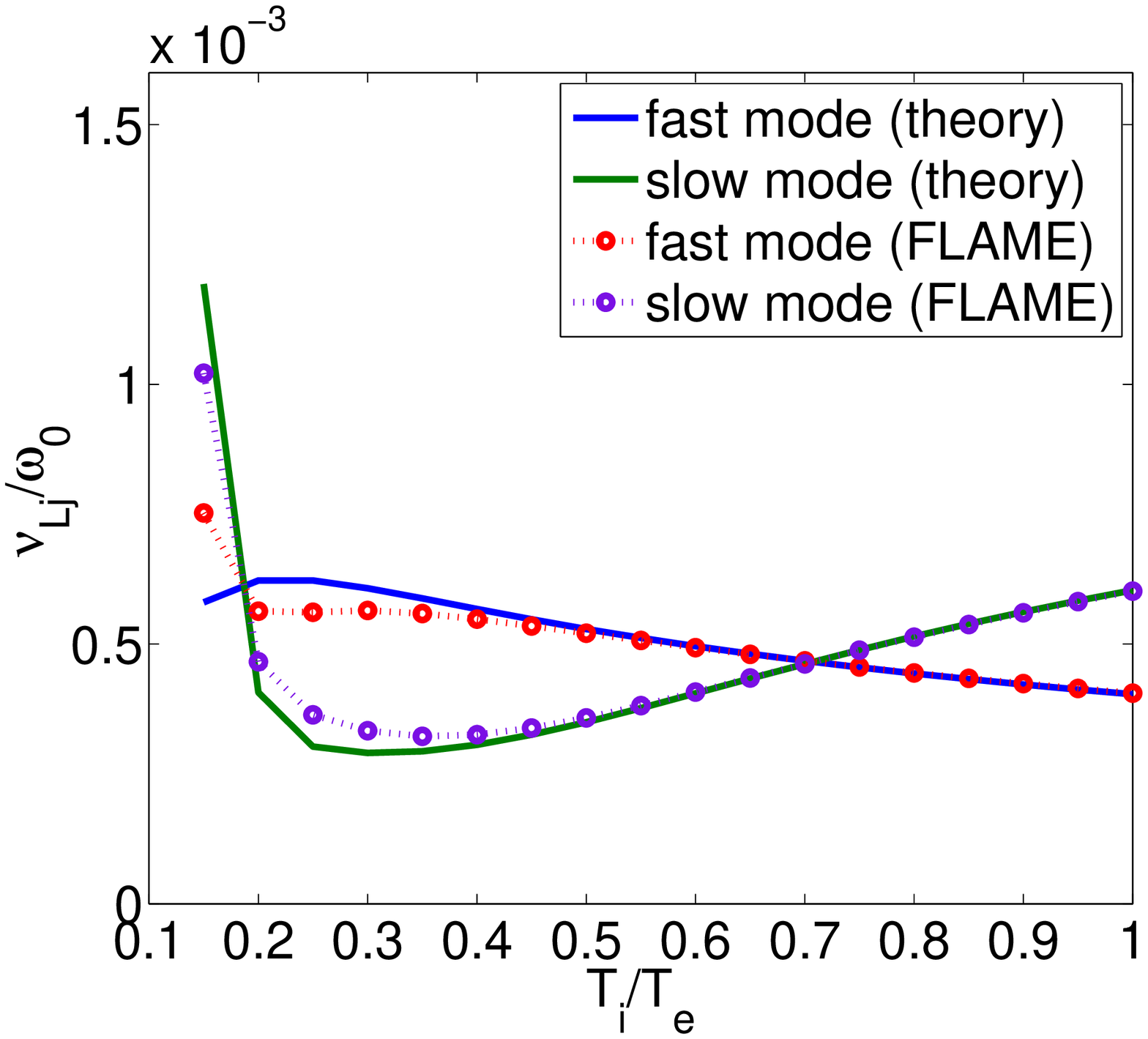}
\caption{(Color online) Effective Landau damping rates for two modes
of the weekly damped ion-acoustic wave in {\it FLAME} code with (a)
$T_e=1.6$keV, $n_e=0.2n_c$, $k\lambda_D=0.224$ and (b) $T_e=3.5$keV,
$n_e=0.2n_c$, $k\lambda_D=0.331$.} \label{fig_landau}
\end{figure*}

In the Landau damping solver,  Eqs. (\ref{e26}) and (\ref{e30}) were
solved in the spectral ($k$-)space using a real-space moving window.
Fast Fourier Transform (FFT) was performed in the window to
calculate Landau damping rates of the electron plasma waves and the
ion-acoustic waves according to the average density and temperature
in the window using Eqs. (\ref{e31}) and (\ref{e32}). Then inverse
FFT was performed to obtain the damped value of $u_{Lx}$ and
$u_{jx}$ at this point in the real space. In the current 1D version,
the Landau damping rates were
\begin{eqnarray}\label{e31}
{\nu _{Le}}(k) = \sqrt {\frac{\pi}{8}} \frac{\omega
_{pe}}{{(k{\lambda _D})^3}}\exp \left[ { - \frac{1}{2{(k{\lambda
_D})}^2} - \frac{3}{2}} \right],
\end{eqnarray}
for the electron plasma wave \cite{Landau46} and
\begin{eqnarray}\label{e32}
&&\nu _{Lj}(k) = \sqrt {\frac{\pi}{8}} \omega_j(k)
\left(\frac{Z_j}{m_j} \right)^{1/2}\\\nonumber &&+\sqrt
{\frac{\pi}{8}}\omega_j(k)\left(\frac{Z_j T_e}{T_j}\right)^{3/2}
\exp\left[-\frac{Z_j T_e}{2T_j(1 + k^2 \lambda _{D}^2)}-
\frac{3}{2}\right]
\end{eqnarray}
for the single ion species case \cite{Fried61}, where
$\omega_j=k\sqrt{{Z_j}{T_e}/{m_j}(1 + k^2\lambda_D^2)+3T_j/m_j}$ and
$\lambda_D$ is the electron Debye length. For cases with two ion
species, analytic formula of Landau damping rates for different
modes of the weakly damped ion-acoustic waves \cite{vu94} are
available. Since we solved the equations of different ion species,
not different modes of the ion-acoustic wave, in our model, we had
to choose Landau damping rates for different ion species
approximately. Currently, in {\it FLAME}, the analytic formula of
Landau damping rate for the slow mode \cite{vu94} was used as the
Landau damping rate of the heavy ion species and the analytic
formula of Landau damping rate for the fast mode \cite{vu94} was
used as the Landau damping rate of the light ion species. By solving
the eigenvalue equation for the ion-acoustic modes \cite{Williams95}
with these Landau damping rates, we can obtain the effective Landau
damping rates for the fast and slow modes in our fluid model.
Examples of a CH-plasma with fully ionized C and H in $1:1$ ratio
are shown in Figs. \ref{fig_landau}(a) and \ref{fig_landau}(b) for
different $T_e$ and $k\lambda_D$, where the effective Landau damping
rates of two different modes were close to their theoretical value
obtained by the analytic formula \cite{vu94}. So the approximation
in Landau damping solver for two ion species was reasonable for most
of temperature conditions.

\section{Benchmark of 1D FLAME Code}

The 1D \emph{FLAME} code was benchmarked by simulating the
reflectivity of single SRS and single SBS modes under different
laser intensities in a uniform hydrogen plasma in the heavily damped
regime. The plasma parameters were chosen as $n_e=0.1n_c$,
$T_e=3.5$keV, $T_i=1.6$keV, with the incident laser wavelength of
$\lambda_0=0.351\mu$m. The length of the physical domain was
$L=4000c/\omega_0$. In this regime, the reflectivity can be
predicted by a theoretical formula
\begin{eqnarray}\label{e33}
R(1-R)=\varepsilon [{e^{G(1-R)}} -R],
\end{eqnarray} where $G$ is the gain factor of the corresponding
instability \cite{Tang66}. For a single SRS mode,
\begin{eqnarray}\label{e34}
G_R=\frac{\omega_{pe}^2k_L^2v_{os}^2L}{8\omega_R\omega_L\nu_{Le}
v_{gR}},
\end{eqnarray}
and for a single SBS mode,
\begin{eqnarray}\label{e35}
G_B=\frac{Z_jm_e\omega_{pe}^2k_A^2v_{os}^2L}{8m_j\omega_B\omega_A\nu_{Lj}
v_{gB}},
\end{eqnarray}
where $\omega_R$, $\omega_L$, $\omega_B$ and $\omega_A$ are the best
matching frequency of the SRS backscattered light, the Langmuir wave, the SBS
backscattered light and the ion-acoustic wave respectively, $k_L$ and
$k_A$ are the best matching wavenumber of the Langmuir wave and
the ion-acoustic wave, $v_{gR}$ and $v_{gB}$ are the group velocities of
the backscattered light in SRS and SBS respectively, and $v_{os}$ is the
electron quiver velocity \cite{Tang66,Hao13}.

\begin{figure}[htb!]
\includegraphics[height=0.3\textwidth,width=0.35\textwidth,angle=0]{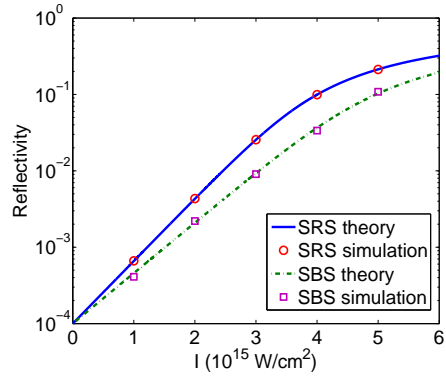}
\caption{(Color online) Reflectivity versus laser intensity in the
benchmark simulations.} \label{fig_benchmark}
\end{figure}

In this regime, Landau damping dominates over collisional damping,
which was neglected in the benchmark. In these benchmark
simulations, we turned off either the ion response part or the
electron response part to simulate pure SRS or SBS, respectively.
The reflectivity in the simulations was measured at the left
boundary of the physical domain, where the incident laser was
launched, using $R={k_s}{A_s}{(0)^2}/{k_0}{A_0}{(0)^2}$ where $k_s$
and $A_s$ denote the wavenumber and amplitude of the backscattered
light, respectively, and $k_0$ and $A_0$ that of the incident light.
Seeds were launched from another antenna at the right boundary with
the resonant frequency of the mode under study. The seed level in
Eq. (\ref{e33}) is defined as $\varepsilon =
{k_s}{A_s}{(L)^2}/{k_0}{A_0}{(0)^2}$, which was chosen to be
$\varepsilon=10^{-4}$ in these simulations. In Fig.
\ref{fig_benchmark}, the solid blue curve and the dash-dot green curve show
the theoretical reflectivities of SRS and SBS, respectively. The red circles
and the purple squares are the relevant simulation results of the 1D {\it
FLAME} code. The good agreement between the simulations and the
theory shows that at least in this regime, the code modeled the
correct physics in exciting and damping of SRS and SBS and also pump
depletion of the incident laser.

\section{The Simulations for LPI in Shock Ignition}
\label{SI}

\begin{figure}[htb!]
\includegraphics[height=0.27\textwidth,width=0.35\textwidth,angle=0]{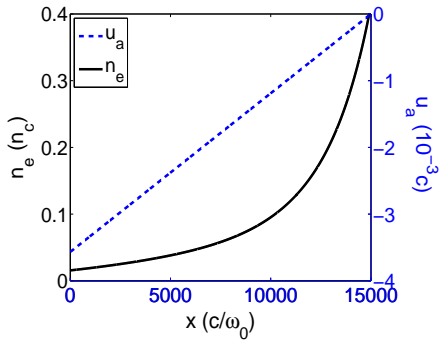}
\caption{(Color online) The normalized density profile (black solid
line) and plasma flow profile (blue dashed line) used in the
simulations.} \label{fig_profile}
\end{figure}

The 1D {\it FLAME} code was used to study SBS and SRS in shock
ignition. These new fluid simulations adopted the same laser and
plasma parameters of the 40+20-beam experiment on OMEGA
\cite{Theobald12} that was simulated in Ref. \cite{Hao16}. Figure
\ref{fig_profile} shows the initial density profile and the plasma
flow profile \cite{Hao16}, fitted from the LILAC \cite{Delettrez}
simulations, in the physical domain of our simulations. The incident
laser had a wavelength of $\lambda_0=0.351\mu m$. The entire
simulation box length was $16000c/\omega_0$, including the length of
the physical domain $L=15000c/\omega_0$ (about $836\mu$m) and two
PML layers of $L_l=500c/\omega_0$ each at the left and right
boundaries. In the physical domain, the density ranged from $0.0156
n_c$ to $0.4 n_c$ with a scale length of $L_n=170 \mu$m near the
$1/4-n_c$ surface. Two ion species were fully ionized C and H in 1:1
ratio. Two sets of typical plasma temperatures were used
\cite{Hao16}. At the launch of the ignition pulse, $T_e=1.6$keV and
$T_C=T_H=0.55$keV, denoted here as the low temperature (LT) case .
At the peak intensity of the ignition pulse, $T_e=3.5$keV and
$T_C=T_H=1.6$keV, denoted here as the high temperature (HT) case.
The grid size was $\Delta x=0.2c/\omega_0$ and the time step was
$\Delta t=0.18/\omega_0$ for both LT and HT cases.

\begin{figure}[htb!]
\includegraphics[height=0.27\textwidth,width=0.35\textwidth,angle=0]{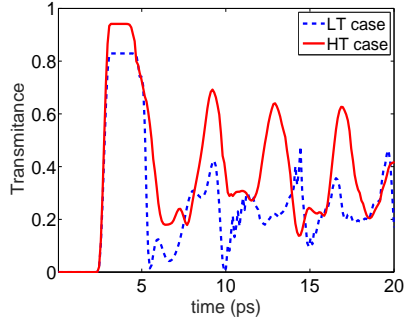}
\caption{(Color online) Transmittance of laser intensity versus
time.} \label{fig_transmittance}
\end{figure}

For the temperature ratio $T_i/T_e\approx 0.34$ in the LT case and
$T_i/T_e\approx 0.46$ in the HT case, the slow mode of the
ion-acoustic wave in the CH plasma has a lower Landau damping rate
than the fast mode as shown in Fig. \ref{fig_landau}. Therefore the
slow mode should be the dominant mode and its damping is higher in
HT than in LT.

\begin{figure*}[htb!]
(a)
\includegraphics[height=0.27\textwidth,width=0.35\textwidth,angle=0]{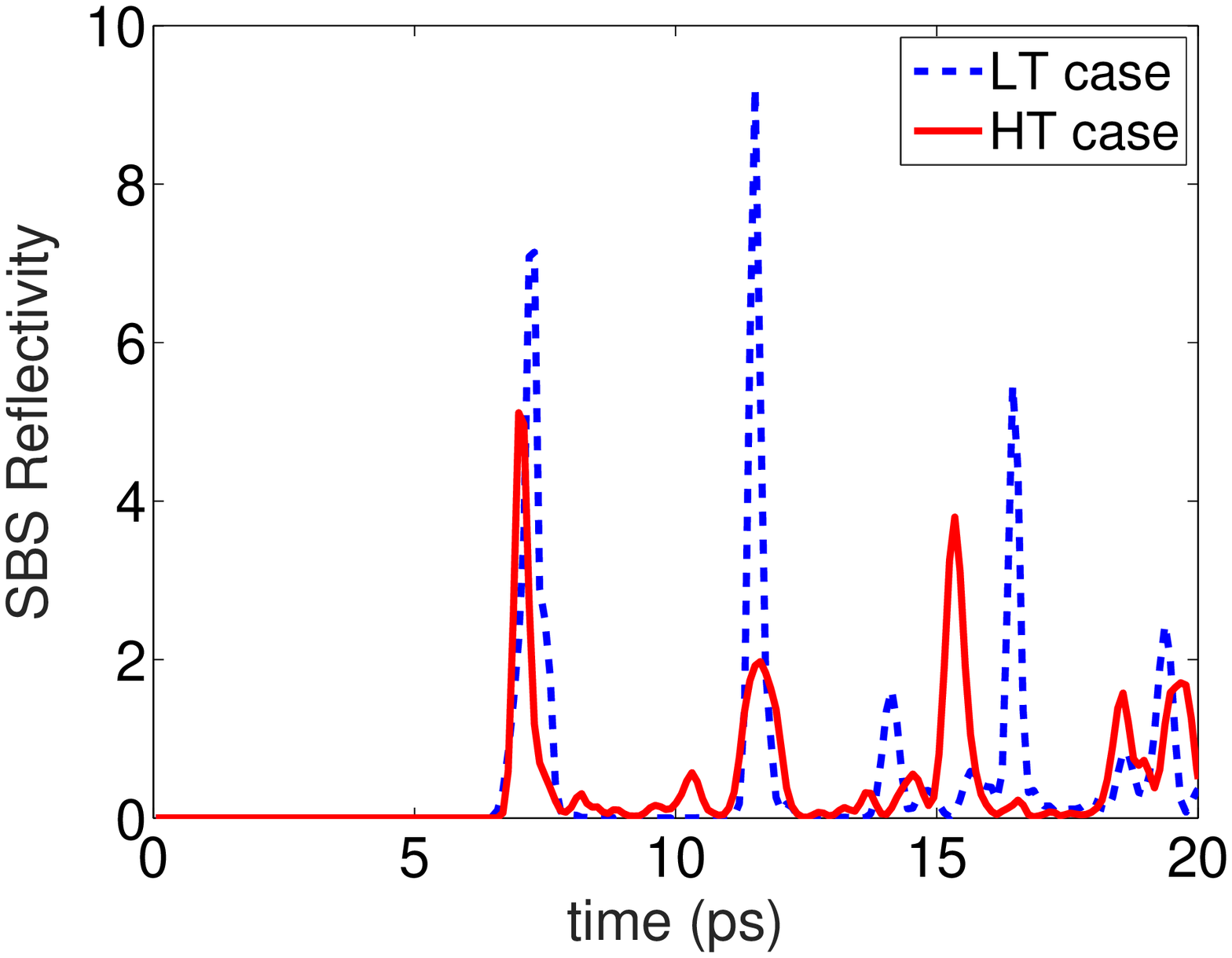}
(b)
\includegraphics[height=0.27\textwidth,width=0.35\textwidth,angle=0]{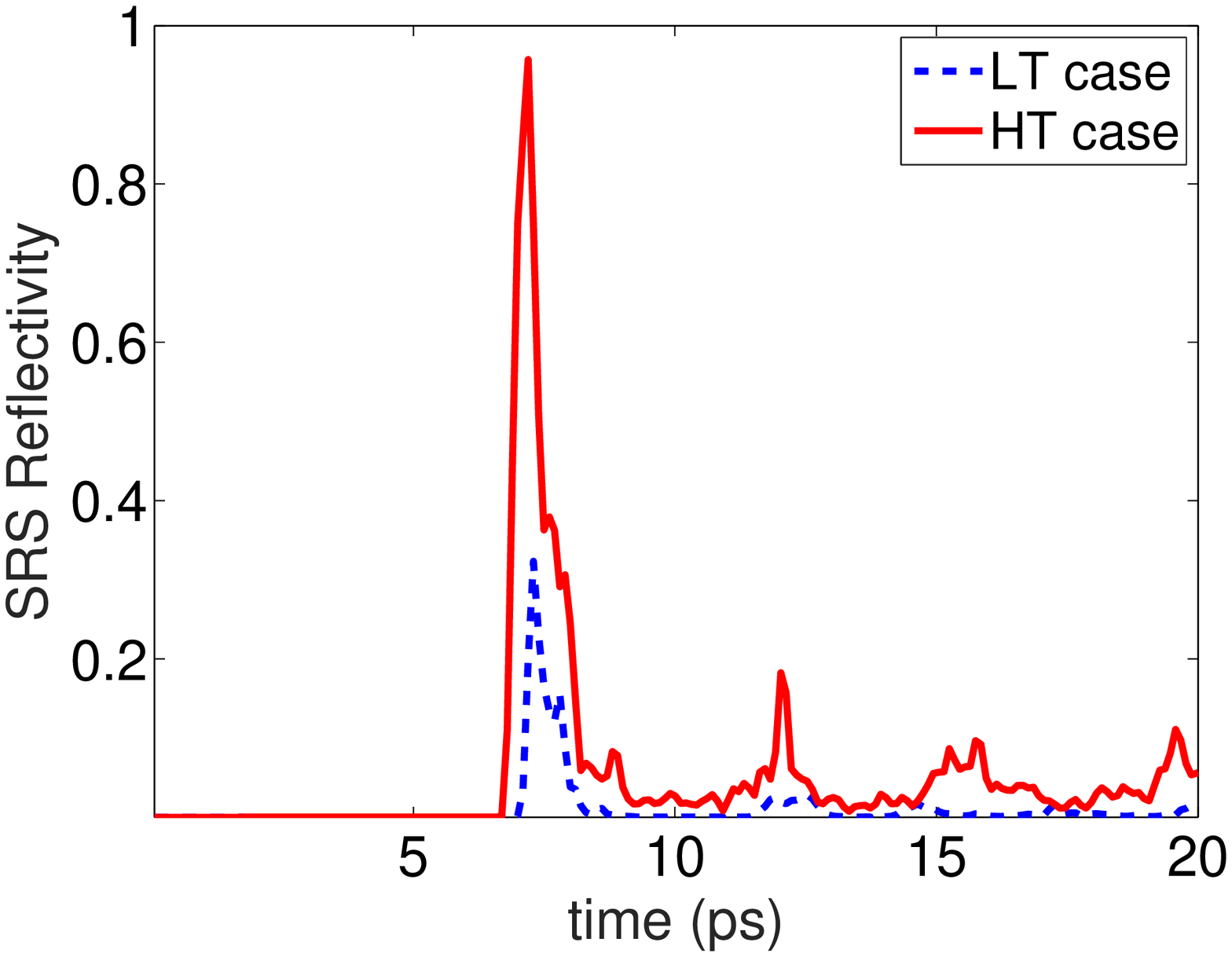}
\caption{(Color online) Reflectivity of (a) SBS and (b) SRS versus
time.} \label{fig_reflectivity}
\end{figure*}
To frequency resolve the incident and backscattered light in {\it
FLAME}, $A_0$ and $A_1$ were dumped every $15$ time steps. Through
2D FFT in time-space windows, the incident and backscattered light
can be separated in the $\omega-k$ phase space. For diagnosis of the
laser transmittance to $n=0.17n_c$, such 2D FFT windows were chosen
in the region of $0.16\sim0.18n_c$ with $2100\Delta x$ in space and
$3000\Delta t$ in time (Fig.\ref{fig_transmittance}). For diagnosis
of the reflectivity, the 2D FFT windows were chosen at the left
boundary of the physical domain with $5000\Delta x$ in space and
$3000\Delta t$ in time (Fig.\ref{fig_reflectivity}).

\begin{figure*}[htb!]
(a)
\includegraphics[height=0.27\textwidth,width=0.35\textwidth,angle=0]{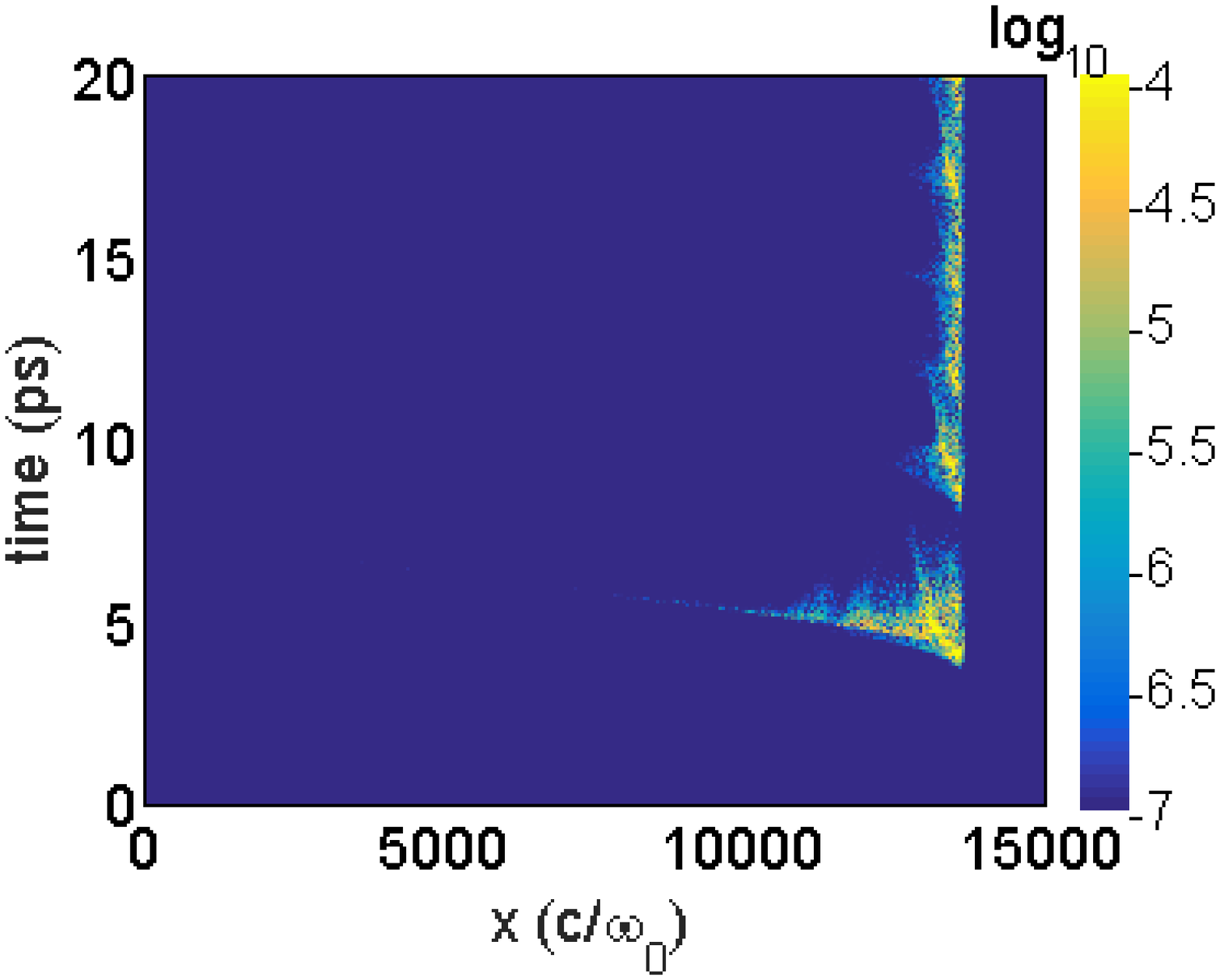}
(c)
\includegraphics[height=0.27\textwidth,width=0.35\textwidth,angle=0]{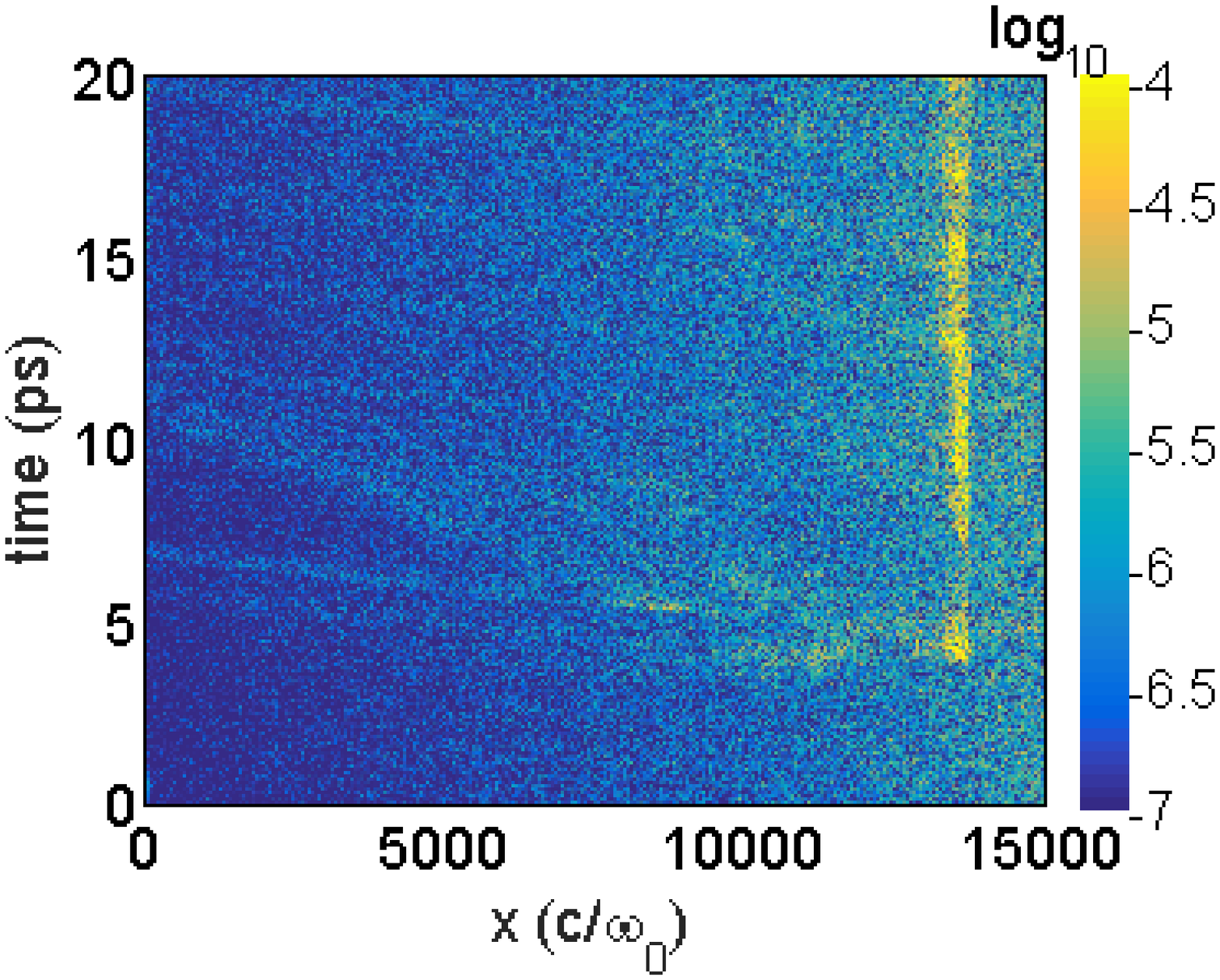}
\\
(b)
\includegraphics[height=0.27\textwidth,width=0.35\textwidth,angle=0]{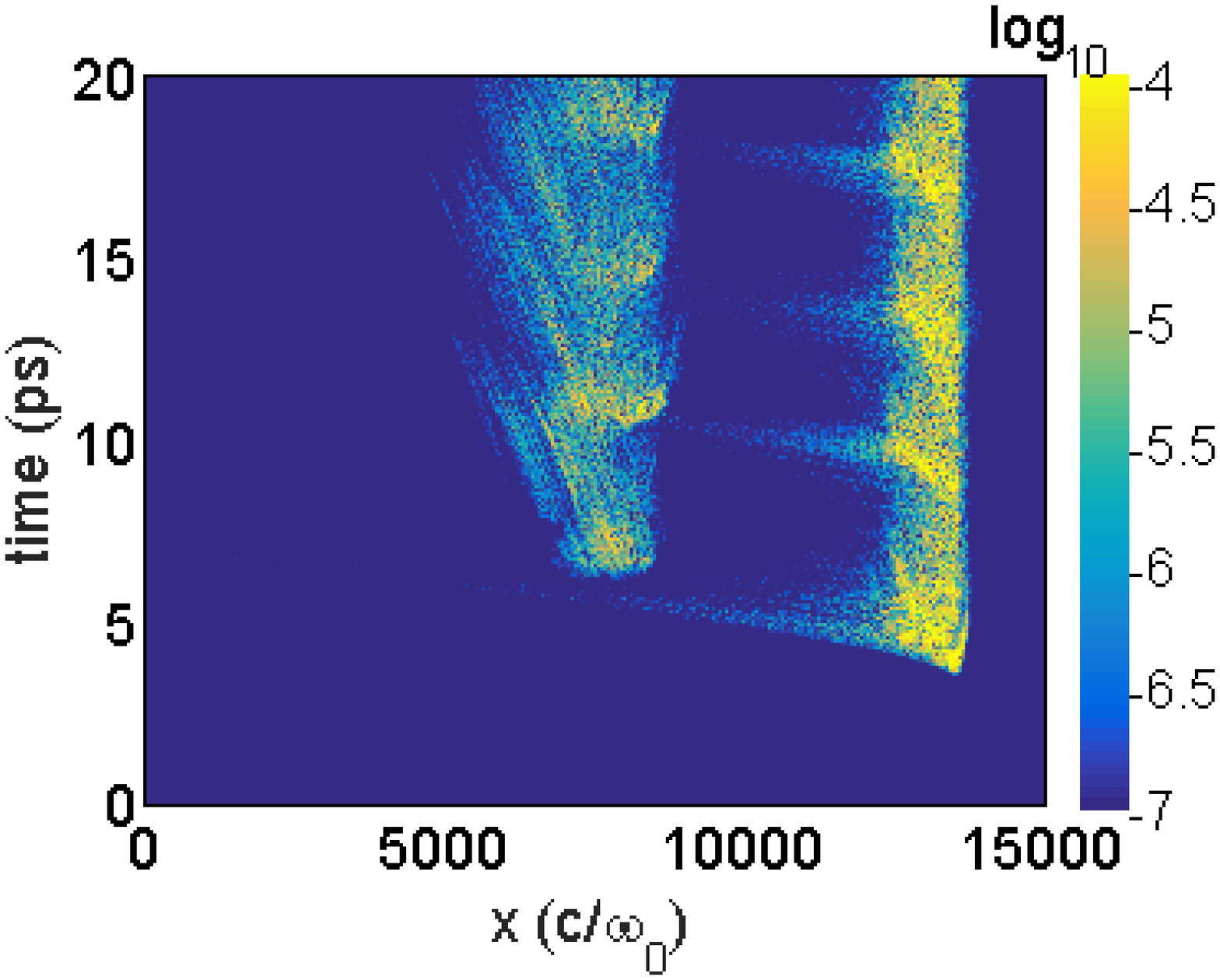}
(d)
\includegraphics[height=0.27\textwidth,width=0.35\textwidth,angle=0]{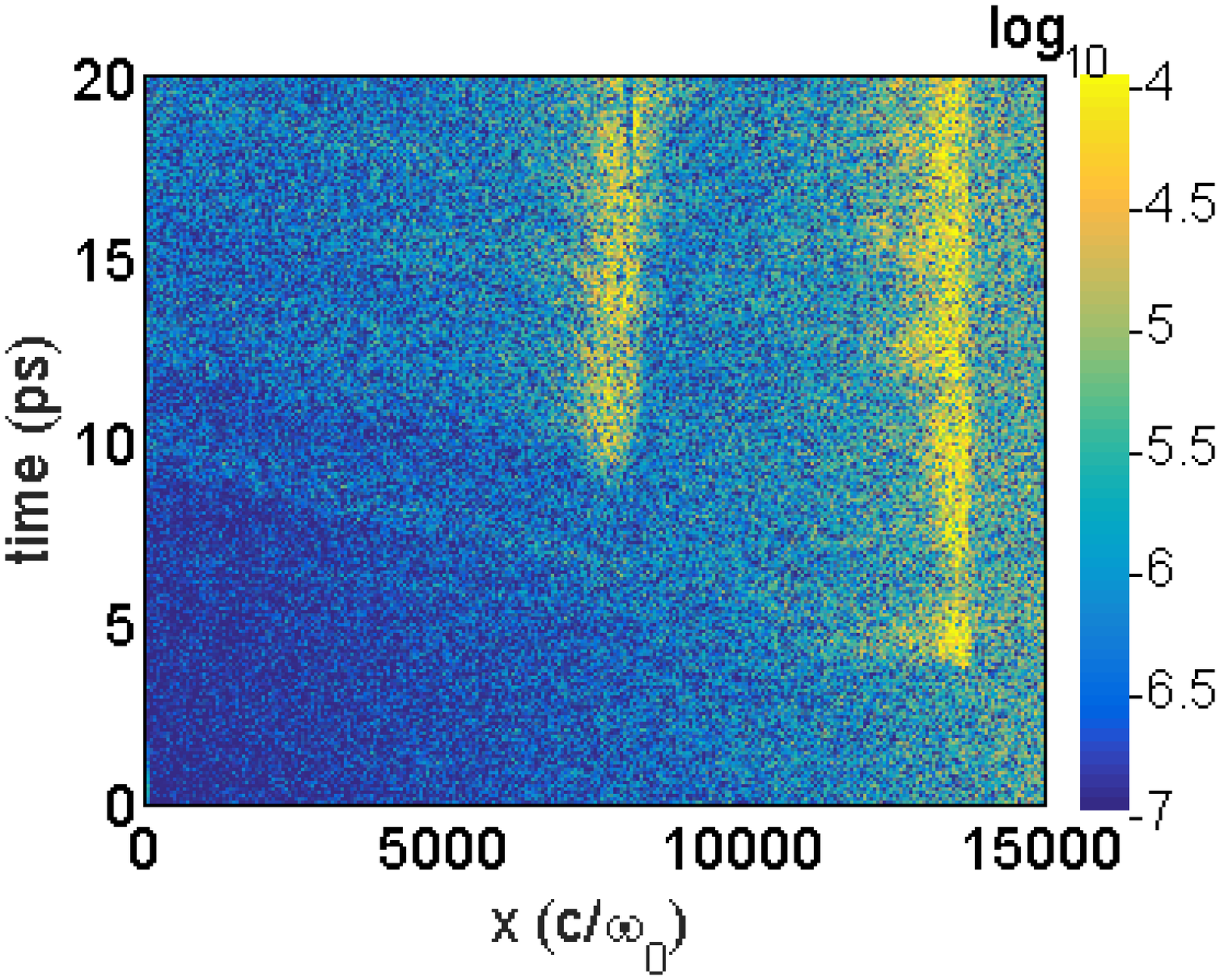}
\caption{(Color online) Evolution of $n_L^2$ in fluid simulations
for the (a) LT and (b) HT cases and evolution of $E_x^2$ in PIC
simulations for the (c) LT and (d) HT cases.} \label{fig_exnl}
\end{figure*}

\begin{table}
 \caption{Average transmittance and reflectivity}
  \begin{center}
    \begin{tabular}{c  c  c  c  c  c  c}
      \hline
      \hline
      & & Temperature & & FLAME & & OSIRIS \\ \hline
      Transmittance & & LT case & & $26\%$ & & $20\%$ \\
        & & HT case & & $38\%$ & & $32\%$ \\
      \hline
      SBS reflectivity & & LT case & & $49\%$ & & $51\%$ \\
        & & HT case & & $41\%$ & & $31\%$ \\
      \hline
      SRS reflectivity & & LT case & & $1.1\%$ & & $13\%$ \\
        & & HT case & & $5.6\%$ & & $25\%$ \\
      \hline
      \hline
    \end{tabular}
  \end{center}\label{table1}
\end{table}

Simulations of both the HT and LT cases were performed for $20$ps
with $I=5\times 10^{15}$W/cm$^2$. Figure \ref{fig_transmittance}
shows the fraction of the incident laser intensity arriving at the
region of $0.16\sim0.18n_c$ as a function of time. Significant pump
depletion started at about $4$ps, and the transmittance was
intermittent and was lower in the LT case. Reflectivities of SBS and
SRS are shown in Figs. \ref{fig_reflectivity}(a) and
\ref{fig_reflectivity}(b), respectively. Bursting SBS reflectivity
had very high instantaneous peaks and was stronger in the LT case,
due to the lower Landau damping rate of the ion-acoustic wave, than
the HT case. The SRS reflectivity was largest in the the first peak
and was significantly smaller than SBS. Therefore SBS was the main
cause for the significant pump depletion. The temporal-averaged
transmittance and reflectivities are listed in Table \ref{table1},
together with the results from OSIRIS \cite{Fonseca02} PIC
simulations of the same conditions.  Both {\it FLAME} and OSIRIS
simulations found similar transmittance and SBS reflectivities in
both the LT and HT cases. They both found lower transmittance,
stronger SBS, and weaker SRS in the LT case than in the HT case. The
two types of simulations showed similar physical trends. One
significant difference is that the {\it FLAME} simulations showed a
much lower SRS reflectivity than the OSIRIS simulations.  The high
SRS reflectivity in the OSIRIS simulation was previously shown
partly due to the strong convective modes in the low density region
that were influenced by the inflated seed levels  \cite{Hao16}.  In
the fluid simulations, we set the magnitude of seeds in {\it FLAME}
code to be about $10^{-9}n_c$, the level of thermal noises
\cite{Berger98}. Without the inflated seed level, SRS was dominated
by the absolute modes.
\begin{figure*}[htb!]
(a)
\includegraphics[height=0.25\textwidth,width=0.35\textwidth,angle=0]{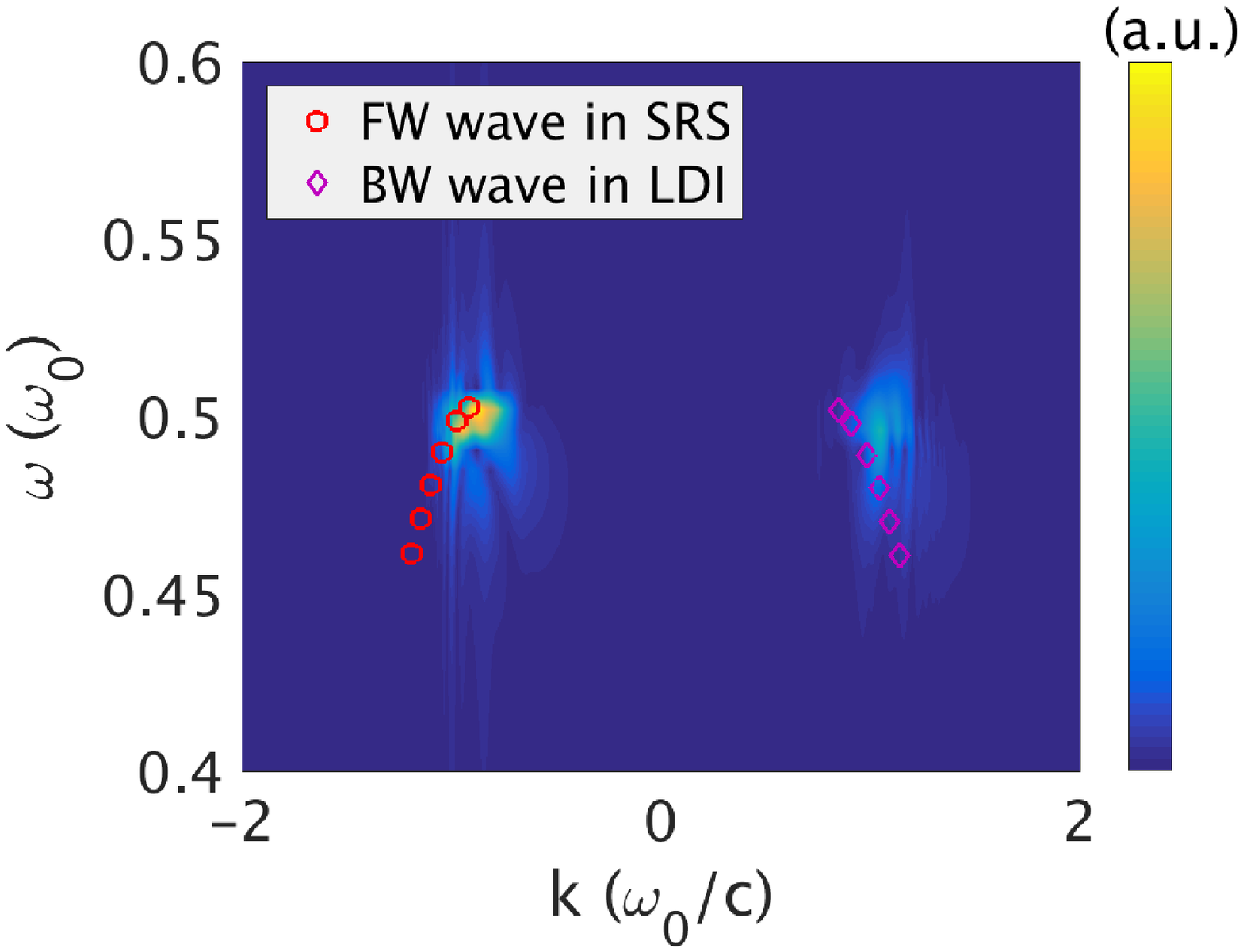}
(b)
\includegraphics[height=0.25\textwidth,width=0.35\textwidth,angle=0]{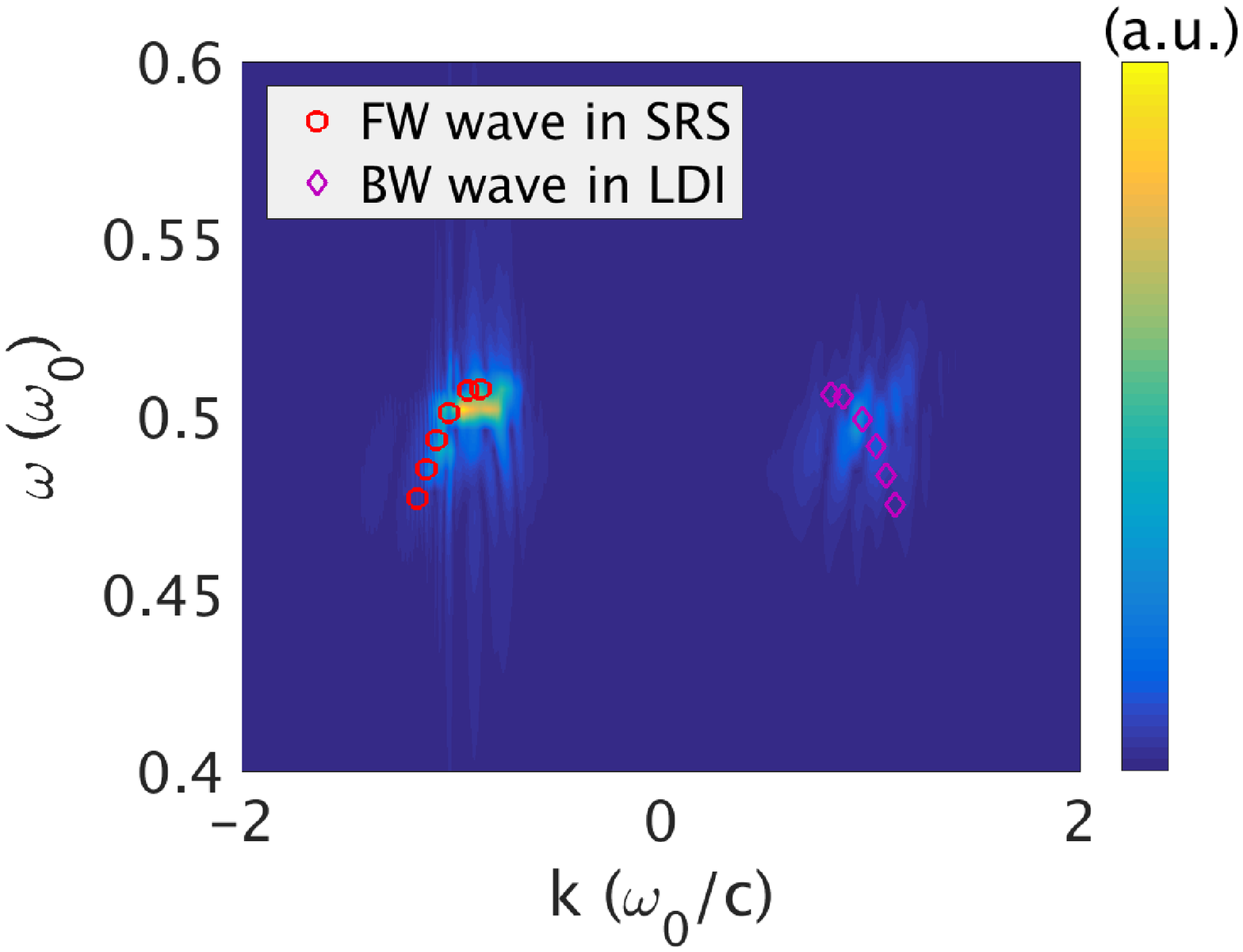}
\caption{(Color online) $\omega-k$ space of $n_L$ near $(1/4)n_c$
for the (a) LT and (b) HT cases.} \label{fig_wknl}
\end{figure*}

\begin{figure*}[htb!]
(a)
\includegraphics[height=0.25\textwidth,width=0.35\textwidth,angle=0]{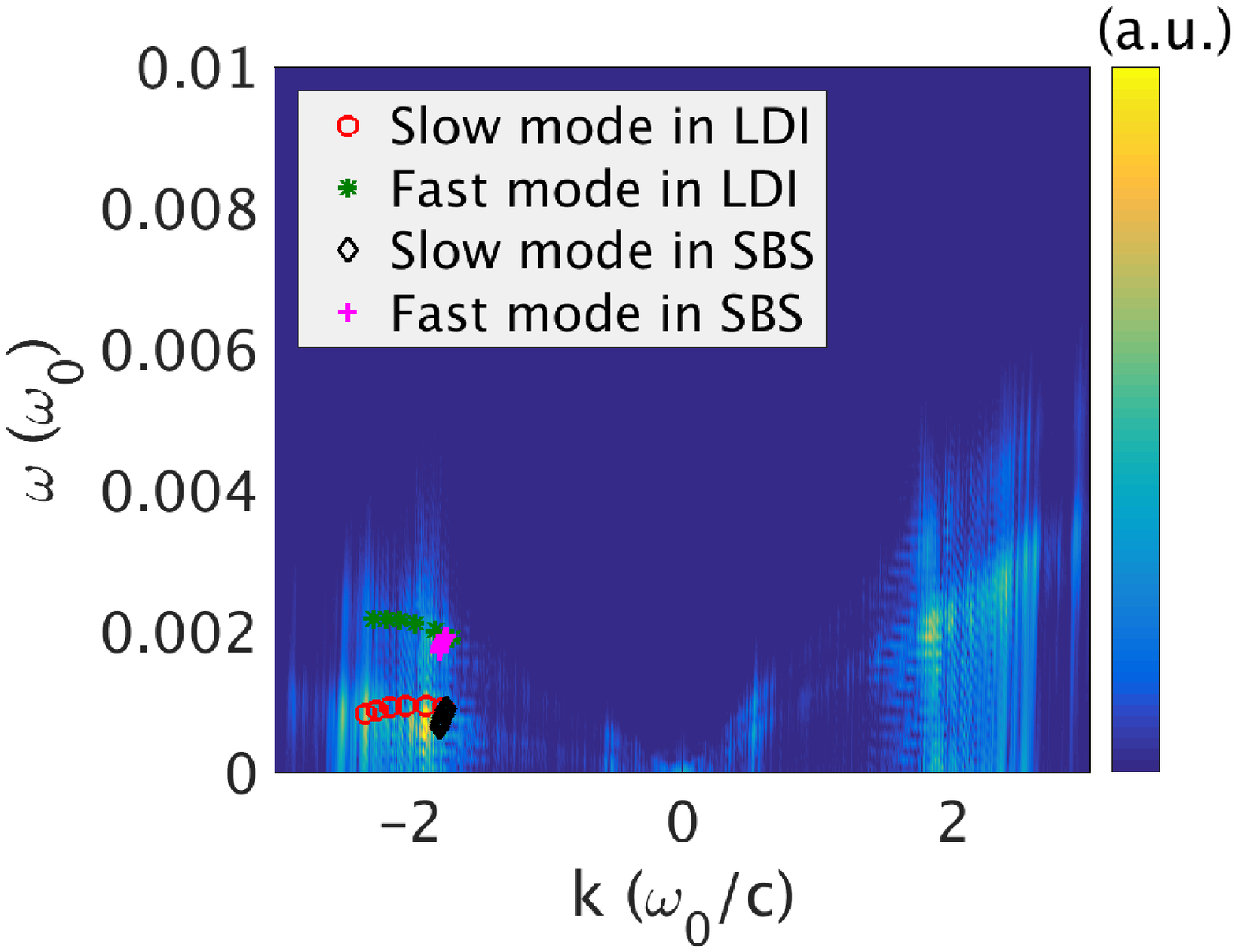}
(c)
\includegraphics[height=0.25\textwidth,width=0.35\textwidth,angle=0]{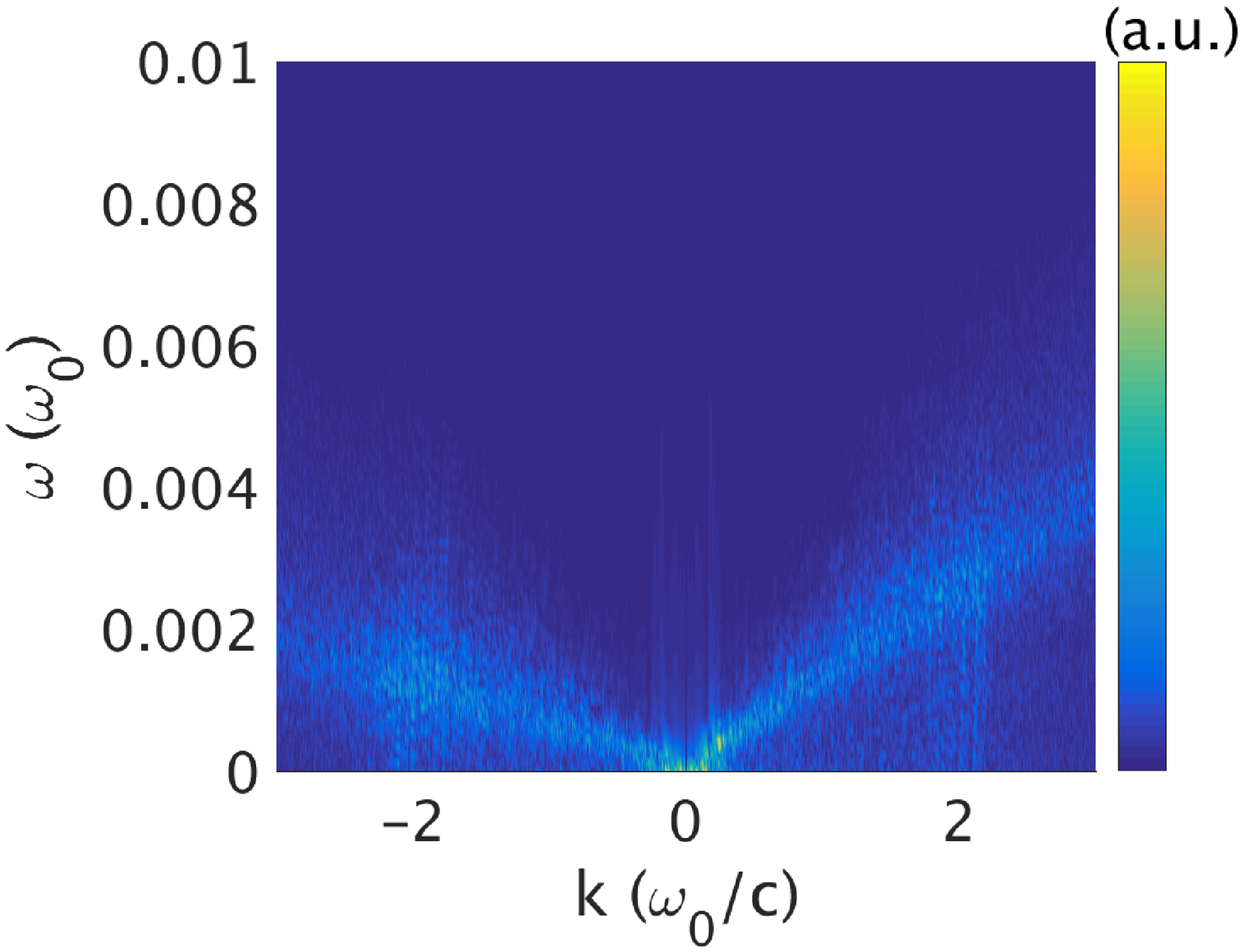}
\\
(b)
\includegraphics[height=0.25\textwidth,width=0.35\textwidth,angle=0]{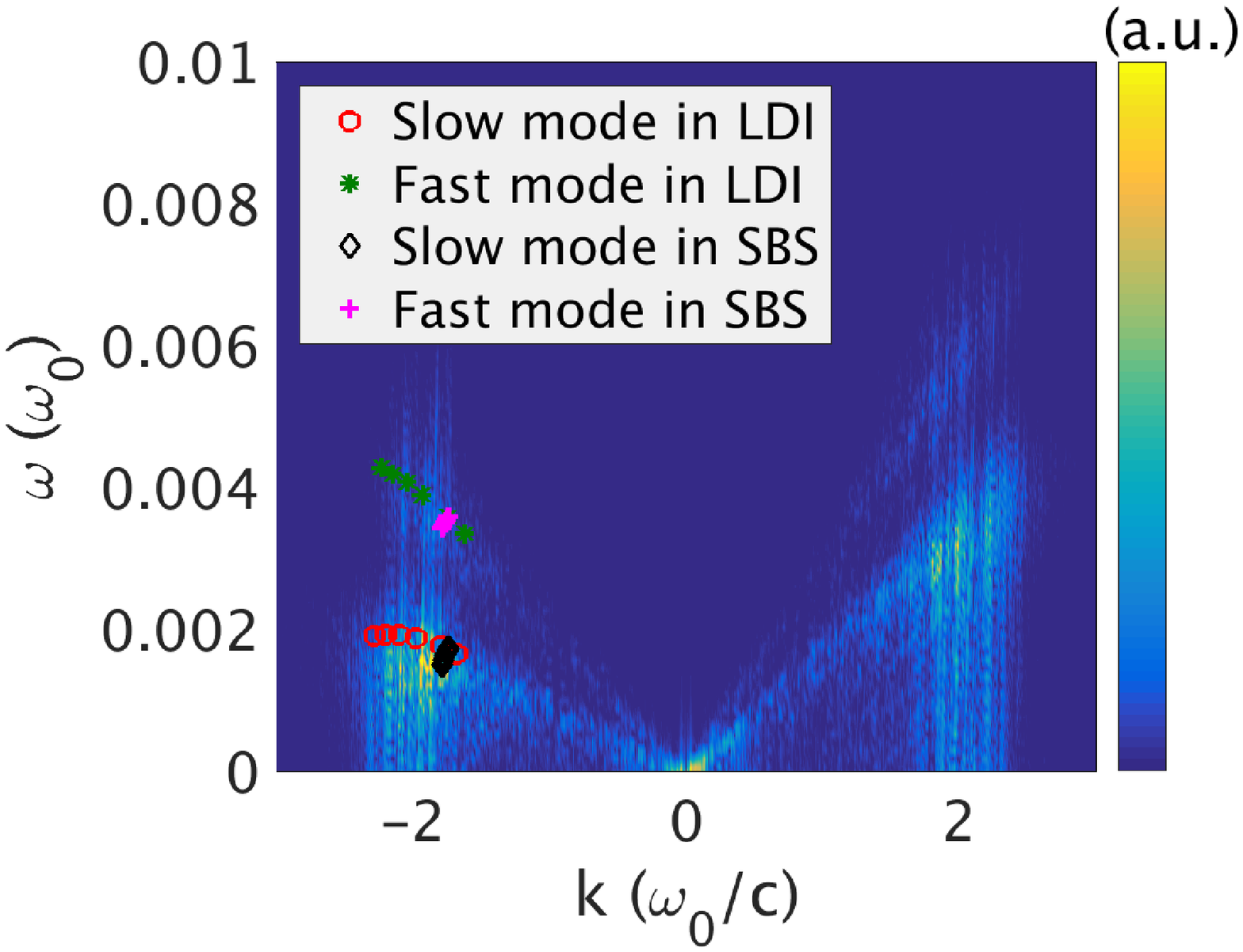}
(d)
\includegraphics[height=0.25\textwidth,width=0.35\textwidth,angle=0]{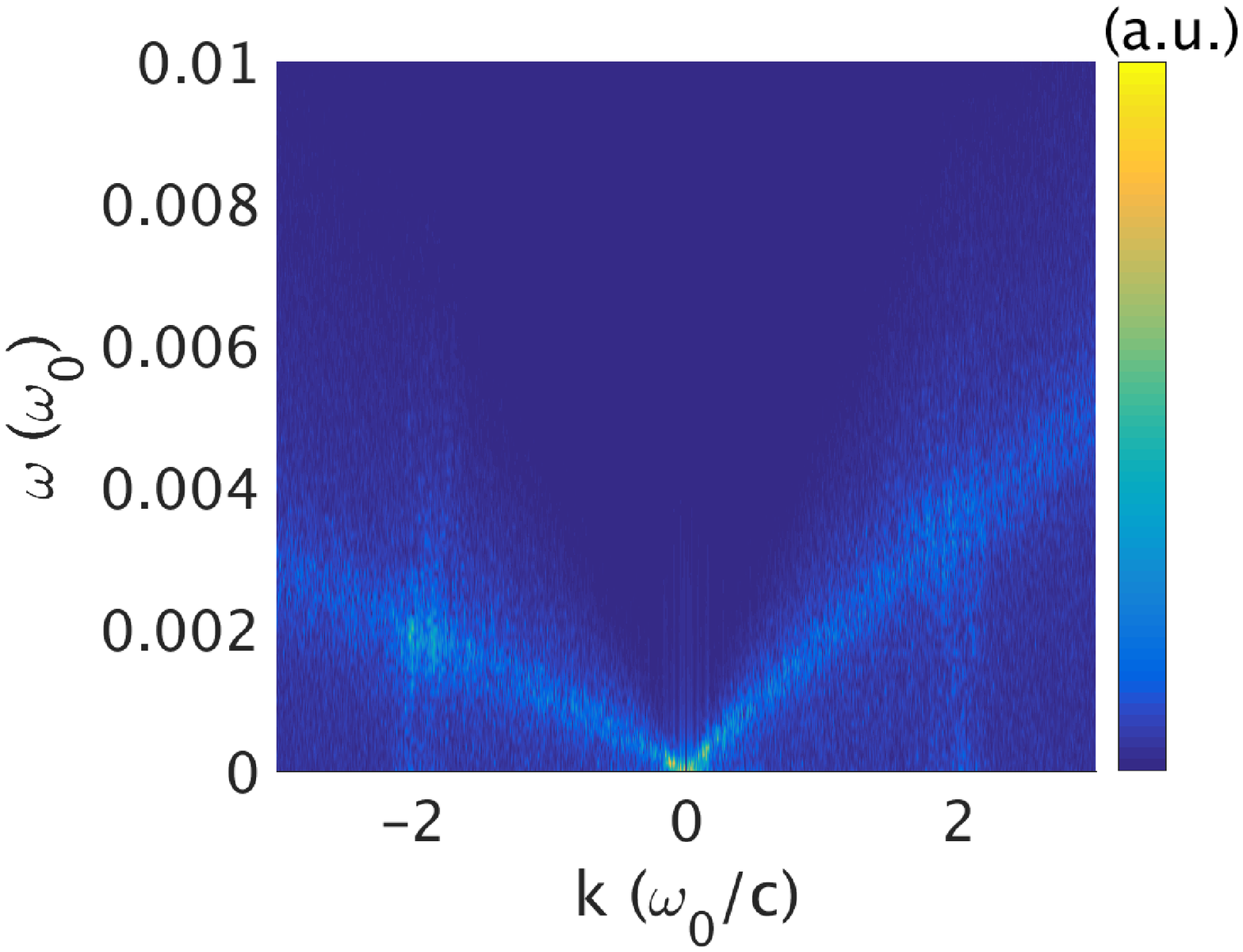}
\caption{(Color online) $\omega-k$ space of perturbations of $n_A$
near $(1/4)n_c$ in fluid simulations for the (a) LT and (b) HT
cases. $\omega-k$ space of perturbations of total charge density of
ions near $(1/4)n_c$ in PIC simulations for the (c) LT and (d) HT
cases.} \label{fig_wkdna}
\end{figure*}

In Figs. \ref{fig_exnl} we compare the plasma wave amplitudes in the
{\it FLAME} and the OSIRIS simulations. In both simulations,
absolute SRS first developed in a narrow region just below
$n=1/4n_c$. Both simulations showed re-scattering of SRS near the
region of $n=1/16n_c$ \cite{Klimo10} in the HT case but not in the
LT. For the re-scattered SRS Langmuir waves near $1/16-n_c$, the
collisional damping dominates over the Landau damping.  The lower
collisional damping in the HT case allowed the re-scattering. The
saturation amplitudes of the plasma waves near the $1/4-n_c$ and the
$1/16-n_c$ surfaces were also comparable in both types of
simulations. All these are evidences that  {\it FLAME} and OSIRIS
modeled key physics of the absolute SRS in similar ways.
Furthermore, the  {\it FLAME} simulations did not show any
significant convective SRS outside the narrow regions near the
$1/4-n_c$ and the $1/16-n_c$ surfaces, unlike the OSIRIS
simulations. We attribute this difference to the inflated seed
levels in the PIC simulations \cite{Hao16}, which we believe was
also the cause of the difference in the total SRS reflectivity in
Table \ref{table1}.

The absolute SRS near $1/4-n_c$ saturated due to pump depletion and
LDI. Figures \ref{fig_wknl}(a) and \ref{fig_wknl}(b) show the
spectra of the electron density associated with the Langmuir waves
$n_L$ for the LT and HT cases respectively. The spectra were
obtained by 2D FFT to $n_L$ in the region of $0.2\sim0.26n_c$ and at
$t=4.2$ps using a window of $4000\Delta x$ and $4500\Delta t$. In
Figs. \ref{fig_wknl}(a) and \ref{fig_wknl}(b), modes in the left
half ($k<0$) were forward (FW) propagating plasma waves, which were
the daughter waves of the absolute SRS.  These modes overlapped with
the red circles representing the theoretical values of the SRS
Langmuir waves calculated from the matching conditions in the
region. And modes in the right half ($k>0$) were backward (BW)
propagating plasma waves, which  overlapped with the pink diamonds
representing the theoretical values of the BW Langmuir waves
calculated from the LDI dispersion relations \cite{Karttunen81}.
These BW waves were the daughter waves of LDI.

Spectra of the ion acoustic waves, from the perturbations of $n_A$,
were also obtained and shown in Figs. \ref{fig_wkdna}(a) and
\ref{fig_wkdna}(b) for the LT and HT cases, respectively. The time
duration of the 2D FFT window was chosen to be $300000\Delta t$. The
forward (FW) ion-acoustic waves in the left half, including both the
fast and slow modes, were from SBS and also from LDI. For our
temperature conditions, the slow mode was the dominant one due to
its smaller Landau damping rate \cite{Williams95}. Both slow and
fast modes overlapped with the theoretical values for the the FW
ion-acoustic wave induced by LDI (the red circles and the green
stars) and by SBS (the black diamonds and the pink plus signs) in
the region of $0.2\sim0.26n_c$. The backward (BW) ion-acoustic waves
in the right half were probably induced by secondary LDI of the BW
Langmuir waves. The phase velocities of the FW and BW ion acoustic
waves were different due to the Doppler shift from the background
flow. These features in the spectra were largely also observed in
the OSIRIS simulations (Figs. \ref{fig_wkdna}(c) and
\ref{fig_wkdna}(d) for the LT and HT cases, respectively).

\begin{figure*}[htb!]
(a)
\includegraphics[height=0.27\textwidth,width=0.35\textwidth,angle=0]{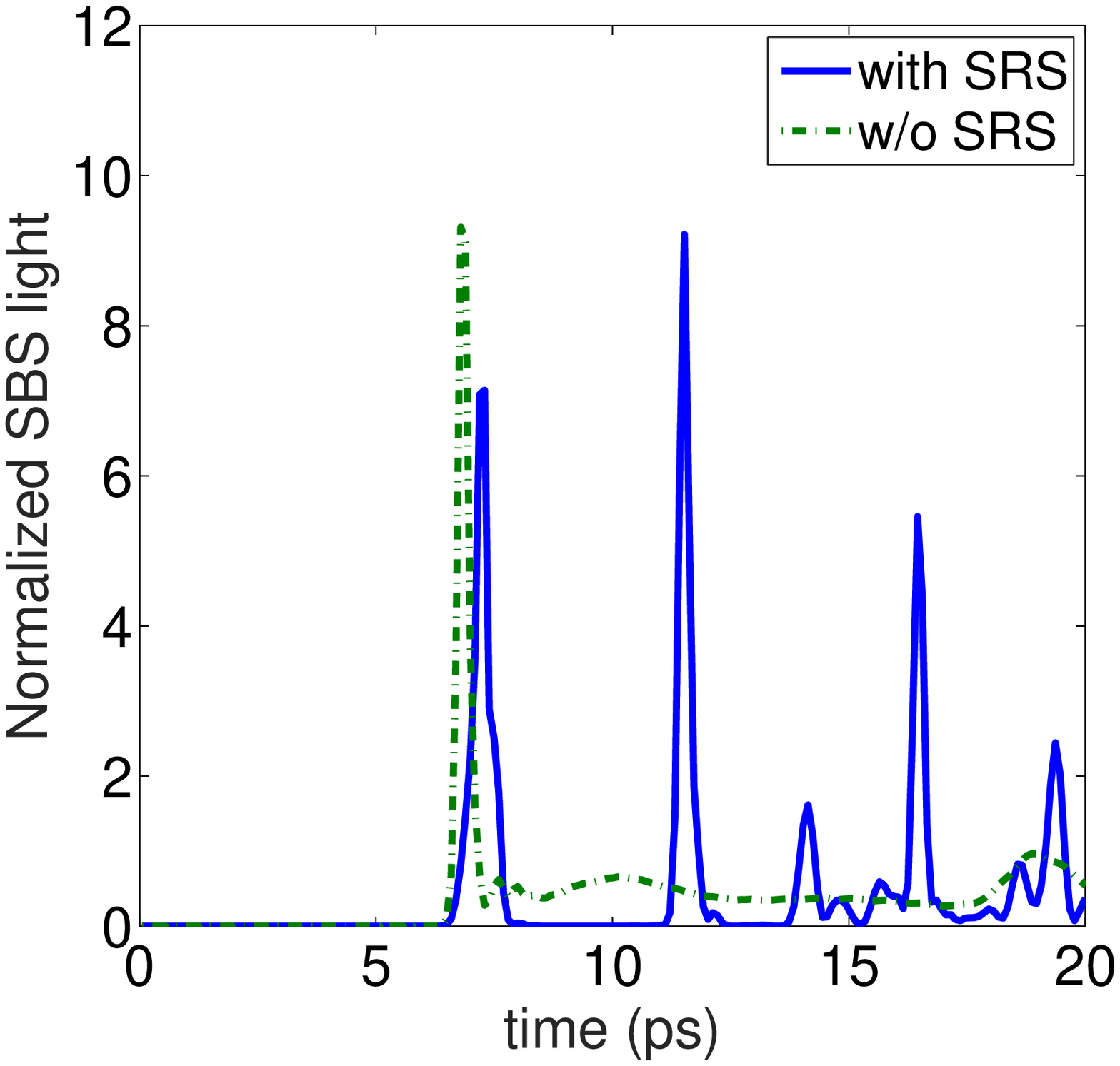}
(b)
\includegraphics[height=0.27\textwidth,width=0.35\textwidth,angle=0]{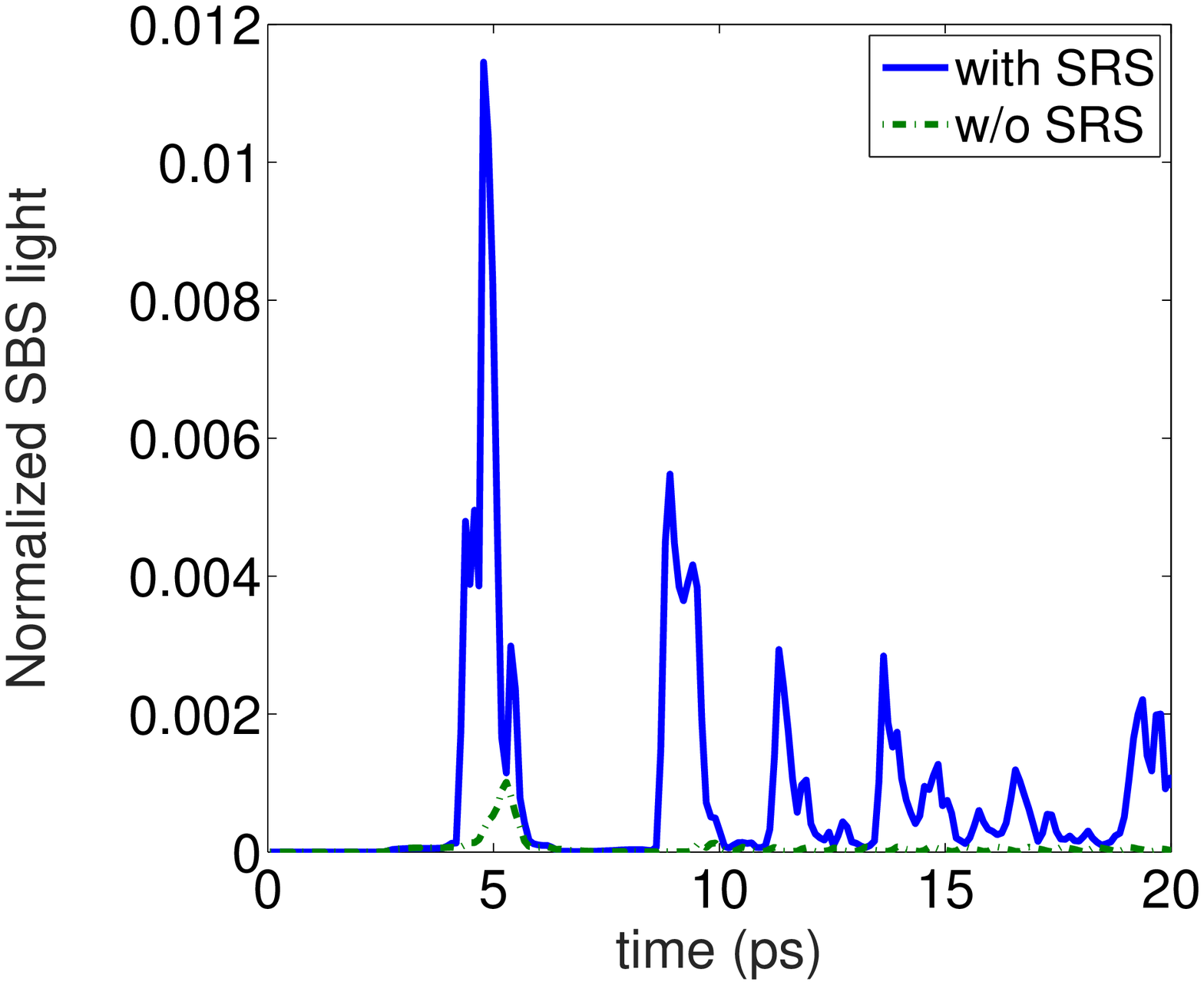}
\caption{(Color online) Comparison of the normalized SBS light intensity (a) at the left
boundary and (b) in the region of
$0.2\sim0.26n_c$ with and without SRS for the LT case.}
\label{fig_compare}
\end{figure*}

Figures. \ref{fig_wkdna}(a) and \ref{fig_wkdna}(b) show that some
modes of the ion-acoustic waves in LDI had similar frequencies and
wavenumbers to the ion-acoustic waves in SBS near $0.25 n_c$. This
indicates that SBS can be seeded by the absolute SRS through LDI
near $0.25 n_c$, rather than by thermal noises. To further
corroborate this, we did a contrasting simulation for the LT case,
where the electron response was turned off to eliminate SRS. Without
competition from SRS, the first peak of the SBS reflectivity was
even slightly stronger (Fig. \ref{fig_compare}a). However after pump
depletion took effect, the SBS reflectivity lost the bursting
pattern and had a time-averaged SBS value of 45\%, which was still
significant and was mainly from the amplification of thermal noises
in the entire box. In contrast, when SRS was present the seed levels
in the $0.2-0.26 n_c$ region were significantly higher (Fig.
\ref{fig_compare}b) and had a bursting pattern. We believed they
were induced by SRS-LDI near $0.25n_c$ and was the cause of the
burst pattern of the SBS reflectivity in Fig. \ref{fig_compare}(a),
which in turn caused the SRS bursts through pump depletion.
Therefore even though the time averaged SBS reflectivities were
similar with and without SRS, their origins were different. Thus we
also conclude that the SBS reflectivity in the PIC simulations were
mostly physical.

\section{Discussion and Summary}
The simulations here and in previous studies
\cite{Riconda11,Weber12,Weber15,Hao16} showed under various
conditions relevant to shock ignition SBS was the dominant
backscatter.  However, experimental results \cite{Theobald12} showed
above $I=4\times10^{15}$W/cm$^2$ SRS reflectivity started to exceed
the SBS reflectivity. The high intensity experimental shots used no
phase plates \cite{Theobald12}. In addition, TPD can compete with
absolute SRS near $0.25n_c$ and affect pump depletion and SBS
seeding. Whether these factors would affect the relative strength in
SBS and SRS reflectivities is an interesting research topic that can
be studied when {\it FLAME} is extended to multi-dimensions.

In summary, we presented a physics model for laser-plasma
instabilities in ICF that is fundamentally multi-dimensional,
multi-fluid and full-wave. This model can be used to study coupling
of major LPI's with controllable noise sources, bridging
envelope-based fluid simulations and full PIC simulations. The
completed 1D version of a nonlinear fluid code {\it FLAME} based on
this model was benchmarked and used to study the coupling of SBS and
SRS for typical parameter conditions in shock ignition. Results
showed strong bursts of SBS in both the low temperature and high
temperature cases, which resulted in strong pump depletion. Absolute
SRS near $0.25 n_c$ saturated by LDI and pump depletion. Part of the
ion-acoustic waves generated in LDI acted as an efficient seed for
SBS. The coupling of SRS and SBS through seeding and pump depletion
caused a bursting pattern in LPI activities. Re-scatter of SRS was
also observed in the high temperature case near $1/16-n_c$. Most of
the simulation results, with the exception of smaller convective SRS
reflectivities due to different noise levels, were consistent with
the PIC simulations. In general, the \emph{FLAME} simulations were
5-10 times faster than the PIC simulations.

\section{acknowledgments}
The authors would like to acknowledge the OSIRIS Consortium for the
use of OSIRIS. This work was supported by DOE under Grant No.
DE-FC02-04ER54789 and DE-SC0012316; and by NSF under Grant No.
PHY-1314734. The research used resources of the
National Energy Research Scientific Computing Center. The support of
DOE does not constitute an endorsement by DOE of the views expressed
in this paper.

Rui Yan is also supported by National Natural Science Foundation of China
(NSFC) under Grant No. 11621202, and No. 11642020; by the Strategic Priority Research Program of the Chinese Academy of Science (Grant No. XDB16); and by Science Challenge Project of China (No. JCKY2016212A505).

\section{Appendix: The physics model}
In this Appendix, we derive the physics model of {\it FLAME}. The
starting point of this model is to group all relevant quantities
according to their frequencies: 1. the ion acoustic wave frequency
and smaller (subscripted $A$); 2. the plasma wave (Langmuir wave)
frequency (subscripted $L$); and 3. the incident laser frequency
(subscripted $0$), with the frequencies satisfying $\omega_A \ll
\omega_L < 0.5\omega_0$ in the underdense laser plasma interaction
regimes. The frequencies of the scattered light, however, usually
overlap with either $\omega_0$ (SBS) or $\omega_L$ (SRS).

Therefore, the relevant quantities can be decomposed as follows:
\begin{equation}
\label{eq:electron:density:decomposition} n_e = n_A+n_L,
\end{equation}
\begin{equation}
\label{eq:electron:phi:decomposition} \phi = \phi_A +
\phi_L+{\phi_0},
\end{equation}
\begin{equation}
\label{eq:electron:vec:decomposition} \vec{u}_e = \vec{u}_A +
\vec{u}_L + \vec{v}_{os},
\end{equation}
where $n_e$ is the electron density, $u_e$ is the electron velocity,
and $\phi$ is the electro-static potential;
$\vec{v}_{os}=\vec{v}_{0}+\vec{v}_{1}$ representing the electron
oscillatory velocities under the incident light ($\vec{v}_{0}$) and
the scattered light ($\vec{v}_{1}$).

This model is also based on the quasi-neutrality assumption,
\begin{equation}
n_A = \Sigma Z_j n_j,
\end{equation}
\begin{equation}
\label{eq:quasineutral:current} n_A \vec{u}_A = \Sigma Z_j n_j
\vec{u}_j,
\end{equation}
where $Z_j$, $n_j$, and $\vec{u}_j$ are the charge number, number
density, and velocity of the $j$th ion species, respectively.

\subsection{The laser propagation}
The Maxwell's equations:
\begin{eqnarray}
\label{eq:faraday}
\nabla\times\vec{E}=-\frac{1}{c}\frac{\partial \vec{B}}{\partial t},\\
\label{eq:ampere}
\nabla\times\vec{B}= \frac{1}{c}\frac{\partial \vec{E}}{\partial t}+\frac{4\pi}{c}\vec{J},\\
\nabla \cdot \vec{E} = 4\pi e (-n_e+\Sigma Z_j n_j) ,\\
\nabla \cdot \vec{B} = 0.
\end{eqnarray}
are convenient to be written in the form of the vector potential
$\vec{A}$ associated with laser field and the electrostatic
potential $\phi$. The electric and magnetic fields are rewritten as
\begin{eqnarray}
\vec{B}=\nabla\times \vec{A},\\
\vec{E}=-\frac{1}{c} \frac{\partial \vec{A}}{\partial t}-\nabla \phi.
\end{eqnarray}

With the Coulomb gauge $\nabla \cdot \vec{A}=0$, Eqs.
(\ref{eq:faraday}) and (\ref{eq:ampere}) become
\begin{equation}
\label{eq:propagation}
\frac{\partial^2 \vec{A}}{\partial t^2}-c^2 \nabla^2 \vec{A} = -c \frac{\partial}{\partial t} \nabla \phi+4\pi c \vec{J}.
\end{equation}

Using $\vec{v}_{0,1}=e\vec{A}_{0,1}/(mc)$, the current
$\vec{J}=(-n_e \vec{u}_e + \Sigma Z_j n_j \vec{u}_j)e $ can also be
decomposed as
\begin{eqnarray}
\vec{J} &=& -en_A(\vec{u}_L +\frac{e\vec{A}_0}{m
c}+\frac{e\vec{A}_1}{m c})\\ \nonumber
&&-en_L(\vec{u}_A+\vec{u}_L+\frac{e\vec{A}_0}{m c} +
\frac{e\vec{A}_1}{m c})\\ \nonumber &=&\vec{J}_0+\vec{J}_1,
\end{eqnarray}
where
\begin{equation}
\vec{J}_0 = -e[n_A \frac{e\vec{A}_0}{m c}+n_L(\vec{u}_L
+\frac{e\vec{A}_1}{m c})],
\end{equation}
\begin{equation}
\vec{J}_1 = -e[n_A(\vec{u}_L +\frac{e\vec{A}_1}{m c})+n_L
(\vec{u}_A+\frac{e\vec{A}_0}{m c})].
\end{equation}
Note that the ion acoustic component $\vec{J}_A$ has been cancelled due to Eq. (\ref{eq:quasineutral:current}).

Then Eq. (\ref{eq:propagation}) can be further decomposed and
simplified as:
\begin{equation}
\label{eq:laser:0} (\frac{\partial^2}{\partial t^2}-c^2 \nabla^2
+\omega_{pe}^2)\vec{A}_0 = - \frac{n_L}{n_A}\omega_{pe}^2 \vec{A}_1
- 4\pi e c n_L \vec{u}_L-{c \frac{\partial}{\partial t} \nabla
\phi_0},
\end{equation}
\begin{equation}
\label{eq:laser:1}
(\frac{\partial^2}{\partial t^2}-c^2 \nabla^2 +\omega_{pe}^2)\vec{A}_1 =
 -c \frac{\partial}{\partial t} \nabla \phi_L - \frac{n_L}{n_A}\omega_{pe}^2 \vec{A}_0 - 4\pi e c (n_A \vec{u}_L + n_L \vec{u}_A),
\end{equation}
where $\omega_{pe}=4\pi n_A e^2/m$ is the local plasma frequency,
$\nabla^2 \phi_L=4\pi en_L$, and ${\partial \nabla \phi_0/\partial
t=-4\pi \vec{J}_0}$, whose purpose is to ensure $\nabla\cdot A_0=0$.

\subsection{The electron evolution}
The fluid equations for the electrons,
\begin{eqnarray}
\label{eq:electrons:mass}
\frac{\partial n_e}{\partial t}+\nabla \cdot (n_e \vec{u}_e)=0,\\
\label{eq:electrons:mom} \frac{\partial \vec{u}_e}{\partial
t}+\vec{u}_e \cdot \nabla \vec{u}_e =-\frac{1}{m n_e} \nabla
p_e-\frac{e}{m}(\vec{E}+\frac{\vec{u}_e\times \vec{B}}{c}),
\end{eqnarray}
can be decomposed using Eqs.
(\ref{eq:electron:density:decomposition},\ref{eq:electron:phi:decomposition},\ref{eq:electron:vec:decomposition}).
The ion acoustic component of the electron mass equation is already
taken care of by the quasi-neutrality conditions. It is straight
forward to obtain the Langmuir time-scale component of Eq.
(\ref{eq:electrons:mass}):
\begin{equation}
\label{eq:mass:e:np} \frac{\partial n_L}{\partial t}+\nabla \cdot
[n_A \vec{u}_L + n_A\frac{e\vec{A}_1}{mc} + n_L \vec{u}_A +
n_L\frac{e\vec{A}_0}{mc}]=0.
\end{equation}

The simplification of the electron momentum equation needs an
equation of state. Here the adiabatic condition for the electrons is
used when handling the Langmuir waves, $[\nabla p_{e}]_L=\gamma T_e
\nabla n_L$ where $\gamma$ is the adiabatic index. Use the identity:
$\vec{a}\cdot \nabla \vec{a}\equiv (\nabla \times
\vec{a})\times\vec{a}+\nabla(|a|^2/2)$, the Langmuir time-scale
terms of Eq.(\ref{eq:electrons:mom}) yield:
\begin{eqnarray}
\frac{\partial \vec{u}_L}{\partial t} &=& \frac{e}{m}\nabla
\phi_L-\nabla[(\frac{e\vec{A}_0}{m c}+\vec{u}_A) \cdot
(\vec{u}_L+\frac{e\vec{A}_1}{m c})]\\ \nonumber && - \gamma v_{th}^2
\frac{\nabla n_L}{n_A} +(\vec{u}_L+\frac{e\vec{A}_1}{m
c})\times\vec{\Omega}+{(\vec{u}_A+\frac{e\vec{A}_0}{m
c})\times(\nabla\times\vec{u}_L)},
\end{eqnarray}
where $v_{th}=\sqrt{T_e/m}$ is the electron thermal velocity,
$\Omega=\nabla \times \vec{u}_A$ is the vorticity of the background
(ion-acoustic time scale) velocity of this plasma.

Neglecting the electron inertia, the slow components of Eq. (\ref{eq:electrons:mom}) become:
\begin{equation}
\label{eq:electron:mom:slow} \frac{\partial \vec{u}_A}{\partial
t}\approx 0 = \frac{e}{m}\nabla
\phi_A-\frac{1}{2}[\nabla(\vec{u}_L+\frac{e\vec{A}}{m c})^2]_A -
\frac{[\nabla p_e]_A}{n_e m},
\end{equation}

which provides the relation between the slow electron static field $\nabla \phi_A$ driving the ions and the quantities in faster time scales.

\subsection{The ion evolution}
Excluding the ion quiver motions, the ion fluid equations are
\begin{eqnarray}
\label{eq:ion:mass}
\frac{\partial n_j}{\partial t}+\nabla \cdot (n_j \vec{u}_j)=0,\\
\label{eq:ion:mom} \frac{\partial \vec{u}_j}{\partial t}+\vec{u}_j
\cdot \nabla \vec{u}_j
=-\frac{1}{M_j n_j} \nabla p_j -\frac{Z_j e}{M_j}\nabla \phi_A,
\end{eqnarray}

Substitute Eq.(\ref{eq:electron:mom:slow}) into Eq.(\ref{eq:ion:mom}), we obtain
\begin{equation}
\frac{\partial \vec{u}_j}{\partial t}+\vec{u}_j \cdot \nabla
\vec{u}_j =-\frac{Z_j m}{M_j}
(\frac{1}{2}[\nabla(\vec{u}_L+\frac{e\vec{A}}{m c})^2]_A +
\frac{[\nabla p_e]_A}{n_e m})-\frac{\nabla p_j}{M_j n_j}.
\end{equation}
Using the iso-thermal condition for the electrons $[\nabla p_e]_A = T_e \nabla n_A$ and the adiabatic condition for the ions $\nabla p_j/p_j=\gamma\nabla n_j/n_j$, we obtain
\begin{equation}
\label{eq:ion:mom2} \frac{\partial \vec{u}_j}{\partial t}+\vec{u}_j
\cdot \nabla \vec{u}_j =-\frac{Z_j m}{M_j}
(\frac{1}{2}[\nabla(\vec{u}_L+\frac{e\vec{A}}{m c})^2]_A) -\frac{Z_j
T_e}{M_j}\frac{\nabla n_A}{n_A}-\frac{\gamma T_j}{M_j}\frac{\nabla
n_j}{n_j}.
\end{equation}
The term $\nabla[\vec{u}_L+e\vec{A}/(m c)]^2$ include components of
multiple frequencies and we aim to only keep the slowest
(ion-acoustic) components and drop the high-frequency parts as much
as possible. So we rewrite Eq.(\ref{eq:ion:mom2}) as:
\begin{equation}
\frac{\partial \vec{u}_j}{\partial t}+\vec{u}_j \cdot \nabla \vec{u}_j\\
=-\frac{Z_j m}{M_j} \frac{1}{2}
\nabla[\vec{u}_L^2+(\frac{e\vec{A}_0}{m c})^2+(\frac{e\vec{A}_1}{m
c})^2+ 2\vec{u}_L \cdot \frac{e\vec{A}_1}{m c}] -\frac{Z_j
T_e}{M_j}\frac{\nabla n_A}{n_A} -\frac{\gamma T_j}{M_j}\frac{\nabla
n_j}{n_j}.
\end{equation}
Note that some high-frequency components still exist in the bracket.
For example, $\vec{u}_L^2$ includes both the $\omega\approx 0$ and
$\omega \approx 2\omega_{pe}$ components. However, the
$2\omega_{pe}$ component is far off resonant in this equation and is
expected not to grow very much.

Finally, we collect all the equations for this model, rewritten as follows:
\begin{enumerate}
\item For the light,
\begin{equation}
(\frac{\partial^2}{\partial t^2}-c^2 \nabla^2
+\omega_{pe}^2)\vec{A}_0 = - \frac{n_L}{n_A}\omega_{pe}^2 \vec{A}_1
- 4\pi e c n_L \vec{u}_L-{c \frac{\partial}{\partial t} \nabla
\phi_0}, \label{first:eq}
\end{equation}
\begin{equation}
(\frac{\partial^2}{\partial t^2}-c^2 \nabla^2 +\omega_{pe}^2)\vec{A}_1 =
 -c \frac{\partial}{\partial t} \nabla \phi_L - \frac{n_L}{n_A}\omega_{pe}^2 \vec{A}_0 - 4\pi e c (n_A \vec{u}_L + n_L \vec{u}_A),
\end{equation}
\item For the electrons,
\begin{equation}
\frac{\partial n_L}{\partial t}+\nabla \cdot [n_A \vec{u}_L +
n_A\frac{e\vec{A}_1}{mc} + n_L \vec{u}_A +
n_L\frac{e\vec{A}_0}{mc}]=0,
\end{equation}
\begin{eqnarray}
\frac{\partial \vec{u}_L}{\partial t} &=& \frac{e}{m}\nabla
\phi_L-\nabla[(\frac{e\vec{A}_0}{m c}+\vec{u}_A) \cdot
(\vec{u}_L+\frac{e\vec{A}_1}{m c})]\\ \nonumber && - \gamma v_{th}^2
\frac{\nabla n_L}{n_A} +(\vec{u}_L+\frac{e\vec{A}_1}{m
c})\times\vec{\Omega}+{(\vec{u}_A+\frac{e\vec{A}_0}{m
c})\times(\nabla\times\vec{u}_L)},
\end{eqnarray}
\item The quasi-neutrality conditions,
\begin{equation}
n_A = \Sigma Z_j n_j,
\end{equation}
\begin{equation}
n_A \vec{u}_A = \Sigma Z_j n_j \vec{u}_j.
\end{equation}
\item For the ions,
\begin{equation}
\frac{\partial n_j}{\partial t}+\nabla \cdot (n_j \vec{u}_j)=0,\\
\end{equation}
\begin{equation}
\frac{\partial \vec{u}_j}{\partial t}+\vec{u}_j \cdot \nabla \vec{u}_j\\
=-\frac{Z_j m}{M_j} \frac{1}{2}
\nabla[\vec{u}_L^2+(\frac{e\vec{A}_0}{m c})^2+(\frac{e\vec{A}_1}{m
c})^2+ 2\vec{u}_L \cdot \frac{e\vec{A}_1}{m c}] -\frac{Z_j
T_e}{M_j}\frac{\nabla n_A}{n_A} -\frac{\gamma T_j}{M_j}\frac{\nabla
n_j}{n_j}. \label{last:eq}
\end{equation}
\end{enumerate}


\end{document}